\documentclass[aps,prfluids,onecolumn,showpacs]{revtex4-1}
\usepackage{amsfonts}
\usepackage{amsmath}
\usepackage{amsthm}
\usepackage{graphics}
\usepackage{graphicx}
\usepackage{subfigure}
\usepackage{amssymb}
\usepackage{graphics}
\usepackage{graphicx}
\usepackage{epstopdf}
\usepackage{color}
\usepackage{subfigure}
\usepackage{pgfplots,tikz}
\usepackage{hyperref}
\newcommand\dertt[1]{ \frac{\partial{ #1}}{\partial t} }

\definecolor{darkcyan}{rgb}{0.0, 0.55, 0.55}

\begin{document}

\title{A matching theory to characterize sound emission during vortex reconnection in quantum fluids}

\author{Davide Proment}
\affiliation{School of Mathematics, University of East Anglia, Norwich Research Park, Norwich NR4 7TJ, United Kingdom}

\author{Giorgio Krstulovic}
\affiliation{Laboratoire J.~L. Lagrange, Observatoire de la C{\^o}te d'Azur, Universit{\'e} de la C{\^o}te d'Azur, CNRS, Nice 06304, France}


\begin{abstract}
In a concurrent work, \emph{Villois et al. 2020} \cite{Villois2020Irreversible} reported the evidence that vortex reconnections in quantum fluids follow an irreversible dynamics, namely vortices separate faster than they approach; such time-asymmetry is explained by using simple conservation arguments. 
In this work we develop further these theoretical considerations and provide a detailed study of the vortex reconnection process for all the possible geometrical configurations of the order parameter (superfluid) wave function. 
By matching the theoretical description of incompressible vortex filaments and the linear theory describing locally vortex reconnections, we determine quantitatively the linear momentum and energy exchanges between the incompressible (vortices) and the compressible (density waves) degrees of freedom of the superfluid. 
We show theoretically and corroborate numerically, why a unidirectional density pulse must be generated after the reconnection process and why only certain reconnecting angles, related to the rates of approach and separations, are allowed.
Finally, some aspects concerning the conservation of centre-line helicity during the reconnection process are discussed.
\end{abstract}
\maketitle

\section{Scientific motivations}

\subsection{Introduction}

Solving the dynamics of the vorticity field is a fundamental problem in fluid mechanics \cite{Saffman:1993aa}.
A fluid possesses in general a continuous three-dimensional vorticity field, but it occurs often in nature that vorticity takes very high values in narrow fluid structures supported in vortex patches like tubes, sheets or filaments.
As such is the case,  studying the dynamics of the vorticity field then results in characterizing the interaction between these structures.
A vortex reconnection is a seminal example of such problem: how do two vortex tubes approach, interact, interchange, and eventually separate?   
Vortex reconnections are an important phenomenon in fluid dynamics. They are observed in the context of plasma physics~\cite{Priest1999}, and both classical~\cite{Kida1994} and quantum fluids~\cite{Fonda25032014,Serafini:2017aa}. Reconnections play an important role transferring energy in superfluid turbulence, fine scale mixing of classical fluid turbulence \cite{Fazle&KarthikPof2011} and solar physics \cite{Zhike2016}. 

Addressing vortex reconnections in (classical) incompressible viscous fluids is an extremely difficult task.
Viscosity is responsible for the diffusion of the vorticity field: an initially finite-supported, or very intense, vorticity tube will gradually loose its distinctiveness in time. Moreover, even for reconnecting processes where the vorticity field of the tubes is so intense that the reconnection time-scale is much faster than viscous diffusion, a variety of new vortex patches like bridges \cite{Kida1994, Yao:2020aa}, ribbons \cite{Pumir:1987aa}, pancakes \cite{Agafontsev:2018aa}, or extra vortex tubes \cite{Kerr:2013aa, McKeown:2020aa} emerge.
Hence, the presence of viscosity is two-folded when studying classical vortex reconnections.
It is a blessing as otherwise, following Kelvin's theorem, the circulation around a vortex filament will simply be transported by the flow and vortex patches will not reconnect; but it is also a curse as it makes the problem mathematically prohibitive.  Furthermore, any description of vortex filament dynamics needs to be accompanied with some modeling of the vortex core. At the moment of reconnection, where the scales at play are of the order of the core size, viscous dissipation becomes extremely strong and it is very difficult to disentangle universal aspects of vortex reconnection from the model-depending physics. What universal aspects of vortex reconnection should prevail in the limit of infinite Reynolds number is an unsolved question. In that sense and from a theoretical point of view, it would be natural to consider the limit where the vortex core is very small. Such limit, naturally arises in quantum fluids.

Quantum fluids, also known as superfluids, are exotic fluids characterized by the complete absence of viscosity.
This (non-classical) property is consequence of their inherent quantum mechanical nature: superfluidity is closely related to Bose--Einstein condensation, a phase transition occurring when a three-dimensional system of bosons is cooled down below a critical temperature or reaches a critical density \cite{Pitaevskii:2016aa}.
In a quantum fluid the vorticity field is filamentary ($ \delta $-supported) \footnote{this is true for mean-field models of quantum fluids, when quantum fluctuations are considered this picture is more complicated.} and because its circulation takes only discrete values, these vortex filaments are called quantized (or quantum) vortices. Quantum vortices are actually topological defects of a macroscopic wave function and they are thus topologically stable non-linear structures. At large distances, they interact each other in the same manner as classical hydrodynamic filaments do. However, unlike classical inviscid vortex filaments, despite the total absence of viscosity, quantized vortex reconnections naturally occur in this system thanks to the presence of dispersive effects \cite{KoplikLevinPRL1993}. The previous reasons make a quantum fluid the ideal system to study the fluid mechanical problem of filamentary vortex reconnections.

In this paper we study superfluid vortex reconnections theoretically and numerically. We provide a theory to explain the origin of the time irreversibly reported in \cite{Villois2020Irreversible} within the framework of the Gross-Pitaevskii model. Our theory explains momentum and energy exchanges during the reconnection process and it is well supported by numerical data.

\subsection{The Gross--Pitaevskii model}

The simplest model that mimics the dynamics of a quantum fluid is probably the Gross--Pitaevskii (GP) equation.
Formally derived for Bose--Einstein condensates made of a dilute gas of bosons \cite{Pitaevskii:2016aa}, it is a mean field nonlinear Schroedinger-type equation that reads
\begin{equation}
i \hbar \dertt{\psi} = -\frac{\hbar^2}{2m} \nabla^2\psi + \frac{4\pi \hbar^2 a_s}{m} |\psi|^2 \psi + V_{\rm ext} \psi \, ,
\label{Eq:GP}
\end{equation}
where $ \psi(\mathbf{r}, t) \in \mathbb{C} $ is the wave-function of the condensate order parameter, $ \hbar $ is the reduced Planck constant, and $ m $ and $ a_s $ are the mass and s-wave scattering length of the bosons, respectively.
In this work, for the sake of simplicity, we assume no external confinement, that is $ V_{\rm ext} \equiv 0 $, and consider the system in a periodic cubic box of volume $ V $.
The following three integrals of motion, corresponding to the total number of bosons, the Hamiltonian (or energy), and linear momentum, respectively, exist
\begin{equation}
N=\int_V |\psi|^2 dV \, ,
\end{equation} 
\begin{equation}
H=\int_V \frac{\hbar^2}{2m}|\nabla\psi|^2 + \frac{2\pi \hbar^2 a_s}{m}|\psi|^4 dV \, ,
\label{eq:H}
\end{equation}
and
\begin{equation}
\mathbf{P}= \frac{i\hbar}{2} \int_V \left( \psi \nabla \psi^\ast - \psi^\ast \nabla \psi \right) dV \, ,
\label{eq:P-GP}
\end{equation} 
where $ (\cdot)^\ast $ stands for complex conjugation.

Introducing the bulk mass density $ \rho_0 = m N / V $, the healing length $ \xi=\sqrt{m/(8\pi a_s \rho_0)} $ and the speed of sound $ c=\sqrt{(4\pi \hbar^2 a_s \rho_0) /m^3} $, Eq.(\ref{Eq:GP}) with $ V_{\rm ext} \equiv 0 $ results in
\begin{equation}
i\dertt{\psi} = \frac{c}{\sqrt{2} \xi} \left(-\xi^2\nabla^2\psi+\frac{m}{\rho_0}|\psi|^2\psi \right) \, .
\label{Eq:GPhydro}
\end{equation}
Written in this manner, the GP equation has the advantage of putting into evidence the time-scale $\xi/c$ and the length-scale $\xi$ at which dispersive effects are important.

The superfluid nature of the system appears evident when one recasts the model in a fluid mechanical framework by using the Madelung transformation
\begin{equation}
\psi({\bf r},t)=\sqrt{\rho({\bf r},t)/m} \exp[i \phi({\bf r},t)/(\sqrt{2}c\xi) ] \,  . 
\label{Eq:Madelung}
\end{equation}
The imaginary and real parts of eq.~(\ref{Eq:GPhydro}) results, respectively, in
\begin{equation}
\begin{split}
& \frac{\partial \rho}{\partial t} + \nabla \cdot \left(\rho \, {\bf v} \right) = 0 \\
& \frac{\partial \bf{v}}{\partial t} + \left( {\bf v} \cdot \nabla \right) \, {\bf v} = -\frac{c^2}{\rho_0}\nabla \rho + c^2\xi^2 \nabla \left( \frac{\nabla^2 \sqrt{\rho}}{\sqrt{\rho}} \right)
\end{split} \, ,
\quad \text{where} \quad {\bf v}({\bf r}, t)=\nabla \phi({\bf r}, t) \, .
\label{Eq:continuity}
\end{equation}
These are nothing but the mass and linear momentum continuity equations for a barotropic, compressible, irrotational, inviscid fluid \cite{nore1997decaying}, that is a superfluid of density $ \rho $ and velocity field $ \mathbf{v} $. 
Despite the irrotational property, vortices arise as filamentary topological defects where $\psi$ vanishes and its argument, see eq. (\ref{Eq:Madelung}), changes by $ n $-multiples of $ 2\pi $.
If so happens, the fluid mechanical circulation evaluated over a closed curve $ \gamma $ around one of these defects is thus quantized and reads
\begin{equation}
\mathcal{C}=\oint_\gamma {\bf v} \cdot d{\bf l} = \sqrt{2}c\xi \oint_\gamma \nabla \arg(\psi) \cdot d{\bf l} = n \Gamma \, , \quad \text{where} \quad n \in \mathbb{Z} \quad \text{and} \quad  \Gamma= 2\pi \sqrt{2}c\xi= \frac{h}{m}
\end{equation}
is the quantum of circulation. This constraint implies that the velocity field ${\bf v}$ diverges as $r^{-1}$, where $r$ is the distance to the vortex. Note that even if $ \arg(\psi) $ is ill-defined at any vortex point, the wave-function $ \psi $ is a regular field. The pretended singularity is just a consequence of the Madelung transformation that is not defined on the topological defects. By definition, the density vanishes on the vortex lines making, for instance, the energy \eqref{eq:H} and the momentum \eqref{eq:P-GP} well defined quantities. The typical size of a vortex core is controlled by the dispersion of the system and it is thus of the order of one healing length $\xi$. Numerically, vortex lines are easily detectable by plotting the iso-surfaces of a low density value compare the bulk, or, if great precision is needed, by tracking accurately the nodal lines themselves \cite{VilloisTrackingAlgo, PhysRevE.93.061103}. 

In a nutshell, if one forgets about density waves, a superfluid described by the GP equation can be considered pictorially as a collection of vortex filaments with a vortex core of size of order $\xi$, whose interaction at large distances is roughly given by the Biot-Savart based vortex filament model \cite{BustamanteNazarenko}. 

\subsection{Irreversible dynamics in quantized vortex reconnections\label{Sec:Irre}}

\begin{figure}
\centering
\includegraphics[width=0.5\columnwidth]{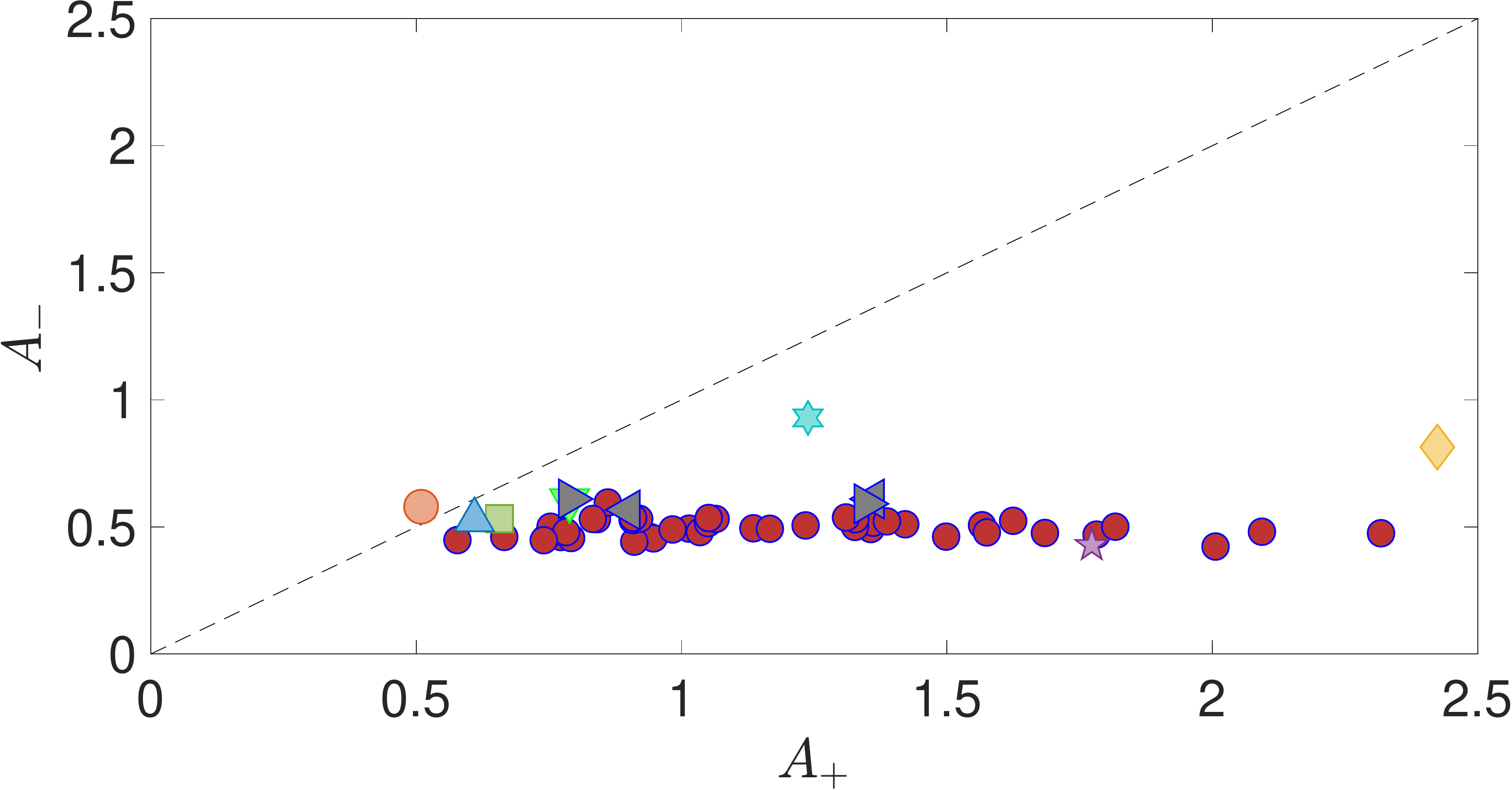}
\caption{(Color online) Values of approaching and separation pre-factors $A^+$ and $A^-$. Red circles correspond to data of \cite{Villois2020Irreversible}. Gray left and right triangles correspond to reconnections of free and trapped vortices respectively, from Galantucci et al.\cite{galantucci2019crossover}. Other symbols are taken from Villois et al. \cite{villoisPRF2018}.\label{Fig:ApAm}}
\end{figure}

Although the possibility of quantum vortex reconnections were suggested by Feynman in the 50's~\cite{Feynman195517}, the first numerical evidence of quantized vortex reconnections, within the GP model, was only given almost 40 years later by Koplik and Levine~\cite{KoplikLevinPRL1993}. Following this seminal work, many numerical and theoretical studies were undertaken in the following decades.  
Many of those works focused on characterizing their rate of approach and separation, on the macroscopic angle between two reconnecting vortex filaments \cite{nazarenko2003analytical, Kursa&Bajer&LipniackiPRB2011, Zuccher&Caliari&Baggaley&BarenghiPof2012, villoisPRF2018, Rorai:2016aa, Serafini:2017aa, Galantucci:2019aa, Rica:2019aa}, on the sound emission following a reconnection event \cite{PhysRevLett.86.1410, Zuccher&Caliari&Baggaley&BarenghiPof2012}, and on the evolution of the length of the filament and superfluid helicity throughout the reconnection process \cite{Scheeler28102014PNASDavide, laing2015conservation, diLeoniHelicity, Salman:2017aa, Zuccher&RiccaPRE2015}.
Despite some initial contradicting results, there is now a general consensus that about a reconnection event
the filaments approach and separate following the law
\begin{equation}
\delta^{\pm}(t) = A^\pm (\Gamma |t-t_{\rm r}|)^{1/2} \, ,
\label{Eq:deltaPM}
\end{equation}
where $ \delta $ is the distance between the filaments, $ t_{\rm r} $ is the time at which the reconnection event takes place, and the signs $ \pm $ stand for what happens before (-) and after (+) the reconnection event.  
Even if the scaling $ \delta \propto |t-t_{\rm r}|^{1/2} $ is universal \cite{villoisPRF2018}, the (dimensionless) pre-factors $ A^\pm $ are not: importantly, one usually finds that $ A^+ \gtrsim A^- $, that is filaments approach slower than they separate.
Figure~\ref{Fig:ApAm} reports a collection of $ (A^+, A^-) $ values obtained in the literature for reconnections of very different nature following:  the decay of Hopf links (red circles)~\cite{Villois2020Irreversible};  interactions between vortex lines and rings in homogeneous and trapped superfluids (triangles)~\cite{Galantucci:2019aa} and regular and random configurations of vortex filaments (all other symbols) \cite{villoisPRF2018}.
These results are a clear evidence of the irreversible dynamics of the reconnection process in quantum fluids. Despite the fact that the GP dynamics is time-reversible and conservative, the data plotted in Fig.\ref{Fig:ApAm} cluster in the region $A^+>A^-$, thus exhibiting a time-asymmetry. In other words, following the experience provided by these data, an educated observer could in principle guess the time direction of a reconnection event.
This fact is clear a manifestation of the irreversibility of the process.

\begin{figure}
\centering
\includegraphics[width=1\textwidth]{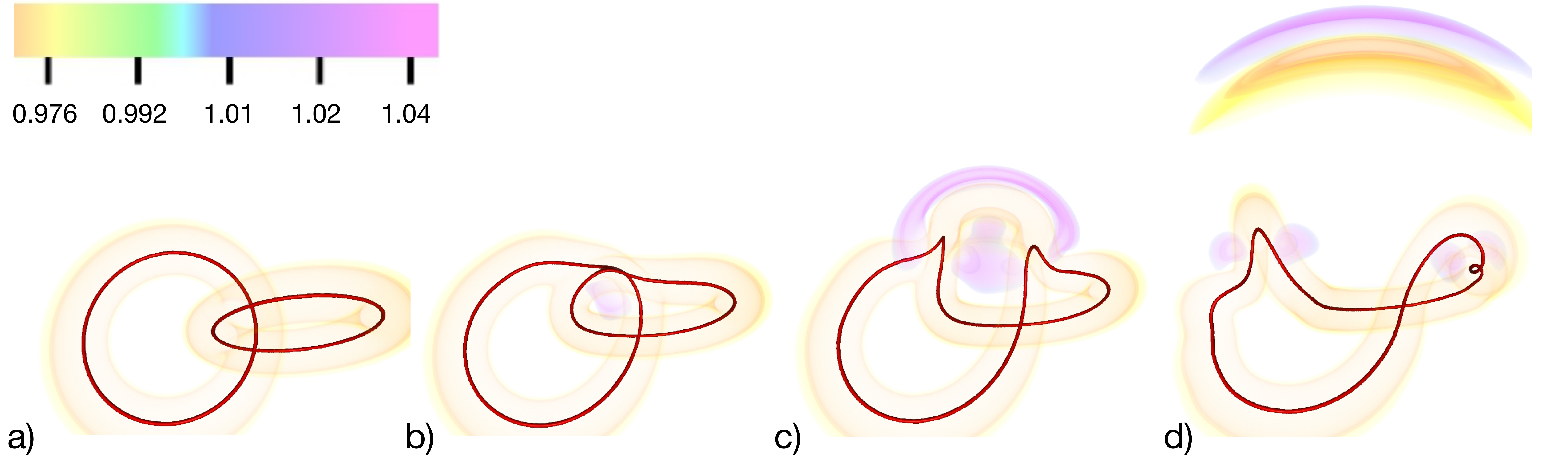}
\caption{(Colour online)
Snapshots of density iso-surfaces of a decaying Hopf link: times correspond to $ t=0 $ \textbf{(a)}, $ t=53 c/\xi $ \textbf{(b)}, $ t=71c/\xi $ \textbf{(c)}, and $ t=101/\xi/c $ \textbf{(d)}.
After the reconnection takes place \textbf{(b)}, a clear sound pulse is created and propagates unidirectionally.}
\label{fig:densityEvol}
\end{figure}

Moreover, as the reconnecting filaments accelerate during a reconnection process, a directional sound pulse is generated in the superfluid.
Figure~\ref{fig:densityEvol} shows a set of snapshots of the evolution of some iso-surfaces of the superfluid density field to highlight the sound emission during a reconnection event following the decay of an Hopf link \cite{Villois2020Irreversible}.
After the reconnection takes place, here $ t_{\rm rec} \simeq 53\xi/c $, a distinctive variation of the bulk density (yellow-to-violet colors) emerges at the reconnection point and propagates non-isotropically.    

Explaining the origin of the asymmetry between the pre-factors $ A^\pm $ and characterizing the directionality and intensity of the sound pulse are the main scopes of this work.
In our reasoning and calculations we use two different limits of the GP model: the linear model, essentially the Schroedinger equation where the nonlinear term of GP is neglected, and the Biot-Savart model, where the compressible degrees of freedom of the superfluid are neglected.   
In a nutshell, our main results follow a simple matching between these two limits and make use of the conservation of total superfluid linear momentum and energy, Eqs (\ref{eq:P-GP}) and (\ref{eq:H}), respectively.

The work is organized as follows.
Section~\ref{sec:geomFil} describes exhaustively all the possible geometrical configurations taken by two reconnecting filaments: subsections~\ref{subsec:recWave} and ~\ref{subsec:recWaveParam} explore the range of parameters of the wave-function about a reconnection event, subsection~\ref{subsec:ParamGeom} relates these wave-function's parameters to a set of geometrical parameters for the vortex filaments, finally subsection~\ref{subsec:paramFil} introduces a useful parameterization of the filaments in terms of the geometrical parameters.
Section~\ref{sec:Deltas} is devoted to the variations in time of the linear momentum and energy of the filaments during a reconnection, and relate these to the emission of the sound pulse: subsection~\ref{subsec:cyl} set the framework for such study, while subsections~\ref{subsec:deltaP} and \ref{subsec:deltaE} contain the detailed calculations of the linear momentum difference and energy variations, respectively.  
Section~\ref{sec:concl} is left for the conclusions and future perspectives.

\section{The geometry of the reconnecting filaments \label{sec:geomFil}}

\subsection{The reconnecting wave-function of the filaments \label{subsec:recWave}}

As previously obtained in \cite{nazarenko2003analytical,villoisPRF2018}, about a reconnection event, the nonlinear term of the Gross-Pitaevskii (GP) model can be neglected, as the superfluid density vanishes at the vortex core. 
The dynamics is thus driven by the linear Schroedinger equation
\begin{equation}
i \partial_t \psi = -\frac{\Gamma}{4\pi}\nabla^2\psi \, .
\label{eq:linGP}
\end{equation}
Without any loss of generality we set the reconnection time $ t_{\rm r}=0 $ and let the reconnection point be the origin of our reference frame. 
The most general second-order-polynomial wave-function initial condition having two nodal lines intersecting at the origin results in
\begin{equation}
\psi_{\rm r}(x, y, z) = \frac{1}{\zeta^{5/2}} \left\{ 
p \left[ z-\frac{A(x \cos\theta+y \sin\theta)^2 + B (-x \cos\theta+y \sin\theta)^2 }{2\zeta} \right] 
+ i \left[ z - \frac{C x^2 + D y^2}{2\zeta} \right] 
\right\} \, ,
\label{eq:genIC}
\end{equation}
where  $ p =\pm 1 $, $ (A, B, C, D) \in \mathbb{R} $ and $ \theta \in [0, \pi] $ are dimensionless parameters, we will call them {\it wave-function parameters} in what follows. $ \zeta>0 $ is a characteristic length.
\begin{figure}
\raggedright (a) \hspace{250pt} (b)  \\
\centering
\includegraphics[width=0.45\textwidth]{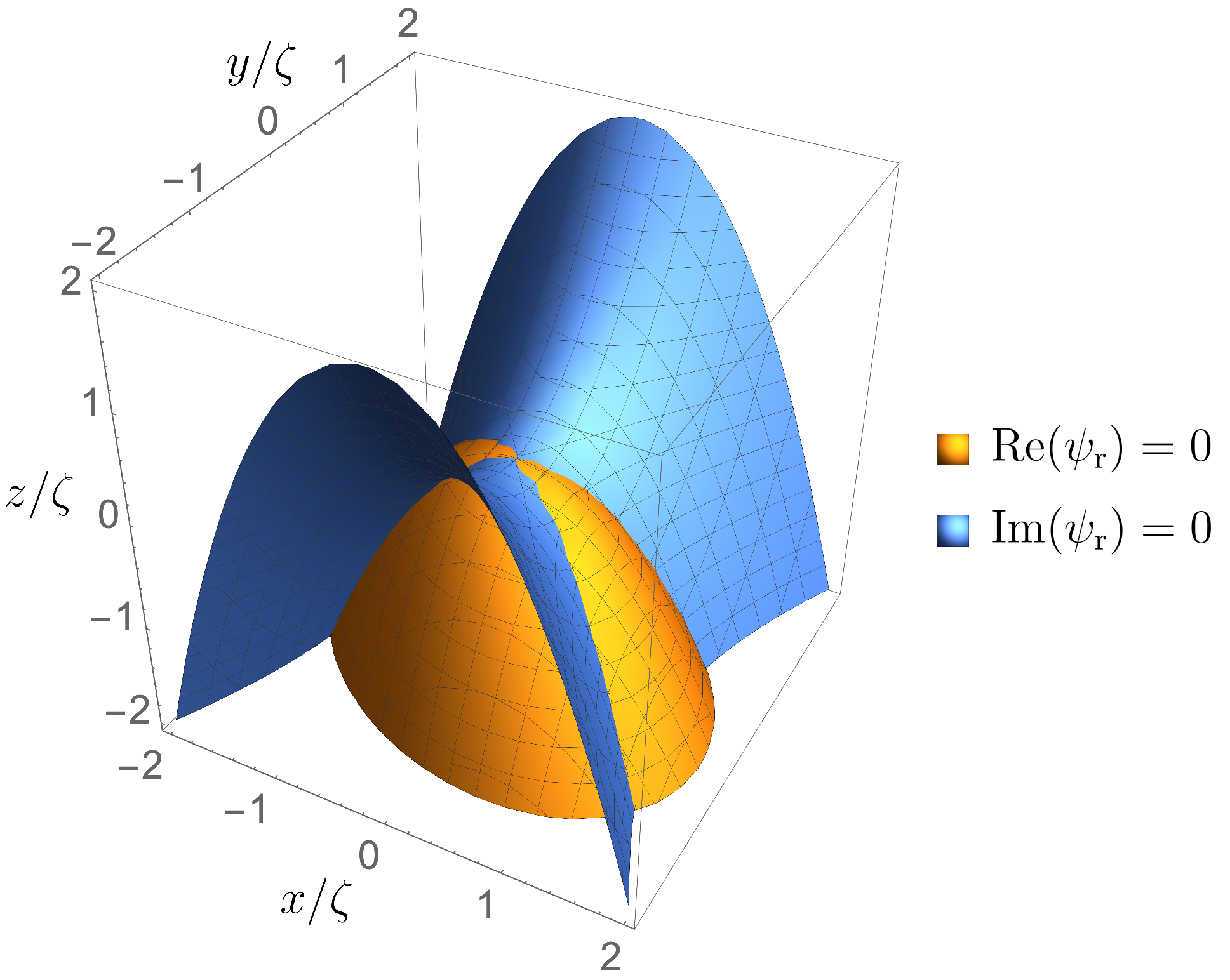}
\quad
\includegraphics[width=0.45\textwidth]{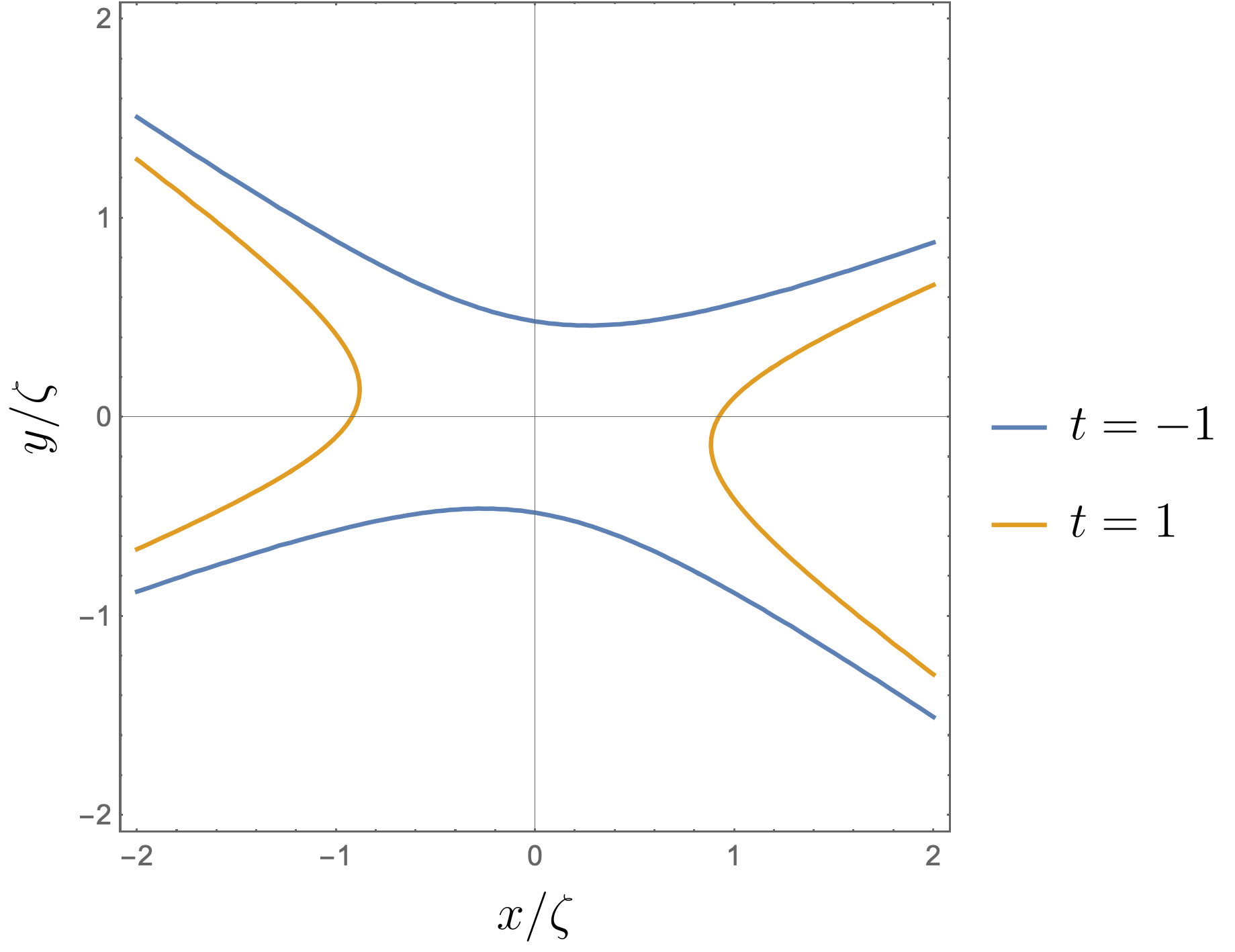}
\caption{(Colour online)
Here $ p=-1, A=-2, B=-1, C=-2, D=1, \theta=\pi/3 $ and lengths are rescaled by $ \zeta $.
\textbf{(a)} An example of the two intersecting iso-surfaces $ \text{Re}(\psi_{\rm r}) =0 $ and $ \text{Im}(\psi_{\rm r}) =0 $ of eq. (\ref{eq:genIC}). 
\textbf{(b)} The hyperbola resulting from the projection of the nodal lines onto the $ z=0 $ plane for the times before, $t=-1$, and after, $ t=1 $, the reconnection time $ t_{\rm r}=0 $.}
\label{fig:genIC}
\end{figure}
The nodal lines of Eq. (\ref{eq:genIC}) can be easily identified by intersecting the two iso-surfaces $ \text{Re}(\psi_{\rm r}) =0 $ and $ \text{Im}(\psi_{\rm r}) =0 $: Figure \ref{fig:genIC}(a) shows an example of this intersection when setting the parameters to $ p=1,  A=-2, B=-1, C=-2, D=1, \theta=\pi/3 $ and where lengths are rescaled by $ \zeta $.

Under the linear Schroedinger operator in Eq.~(\ref{eq:linGP}) the solution in time reads 
\begin{equation}
\psi(x, y, z, t) = e^{it\frac{\Gamma}{4\pi}\nabla^2}\psi_{\rm r}(x, y, z)=\left(1+it\frac{\Gamma}{4\pi}\nabla^2\right)\psi_{\rm r}(x, y, z) \, .
\label{eq:solPsi}
\end{equation}
After some tedious algebra (see the Mathematica notebook available as Supplemental Material \cite{Proment2020MatchingSupplemental}), we obtain that the evolution of the wave-function nodal lines $ \text{Re}(\psi) =0 $ and $ \text{Im}(\psi) =0 $ results in the equations
\begin{equation}
z = \frac{(A+B) (x^2+y^2) + (A-B)  \left[ (x-y)(x+y) \cos(2\theta) + 2xy\sin(2\theta) \right]}{4 \zeta} - \frac{C+D}{4 p \pi \zeta} \, \Gamma t
\label{eq:nodal_re}
\end{equation}
and
\begin{equation}
z = \frac{C x^2 +D y^2}{2 \zeta} + \frac{A+B}{4 p \pi \zeta} \, \Gamma t \, ,
\label{eq:nodal_im}
\end{equation}
respectively.

By simplifying the $ z $-dependence in Eqs. (\ref{eq:nodal_re}) and (\ref{eq:nodal_im}), one finds that the projection of the nodal lines onto the $ z=0 $ plane satisfies the equation
\begin{equation}
(A+B-2C) x^2 + (A+B-2D) y^2 + (A-B) \left[ (x-y)(x+y) \cos(2\theta) + 2 x y \sin(2\theta) \right]= \frac{A+B+C+D}{p \pi} \, \Gamma t \, .
\label{eq:genICProj-xy-theta}
\end{equation}
For a suitable choice of the dimensionless parameters, this relation identifies a hyperbola; an example of it is shown in Fig.~\ref{fig:genIC}(b) where, again, $ p=-1, A=-2, B=-1, C=-2, D=1, \theta=\pi/3 $ and lengths are rescaled by $ \zeta $.
It is important to notice that the $ (x, y) $ axes can always be rotated in order to ensure that the hyperbola asymptotes are mirrored with respect to the two axes.
This property simply reflects the fact that Eq. (\ref{eq:genICProj-xy-theta}) can be re-expressed in its normal form by a suitable rotation that depends on the chosen value of the wave-function parameter $ \theta $. For the sake of simplicity, we set the specific value
\begin{equation} 
\theta \equiv 0. 
\label{eq:condTheta}
\end{equation}
Note however that all the following considerations are actually valid for any value of $\theta$ (see Appendix).
The projection of the nodal lines onto the $ z=0 $ plane now results in
\begin{equation}
-\frac{C-A}{B-D} x^2 + y^2 = \frac{A+B+C+D}{2 (B-D) p \pi} \, \Gamma t \, ,
\label{eq:genICProj-xy}
\end{equation}
by assuming, from now onwards, that $ B \neq D $.

Finally, by simplifying the $ y $-dependence in Eqs. (\ref{eq:nodal_re}) and (\ref{eq:nodal_im}), we find that the projection of the nodal lines onto the $ y=0 $ plane satisfies the equation
\begin{equation}
z =  \frac{BC-AD}{2(B-D) \zeta} \, x^2 + \frac{D(C+D)+B(A+B)}{4(B-D) p \pi \zeta} \, \Gamma t \, ,
\label{eq:genICProj-xz}
\end{equation}
that is a parabola that shifts along the $ z $-axis at constant speed.

\subsection{Region of validity of the other wave-function parameters \label{subsec:recWaveParam}}
\begin{figure}
\includegraphics[width=0.4\textwidth]{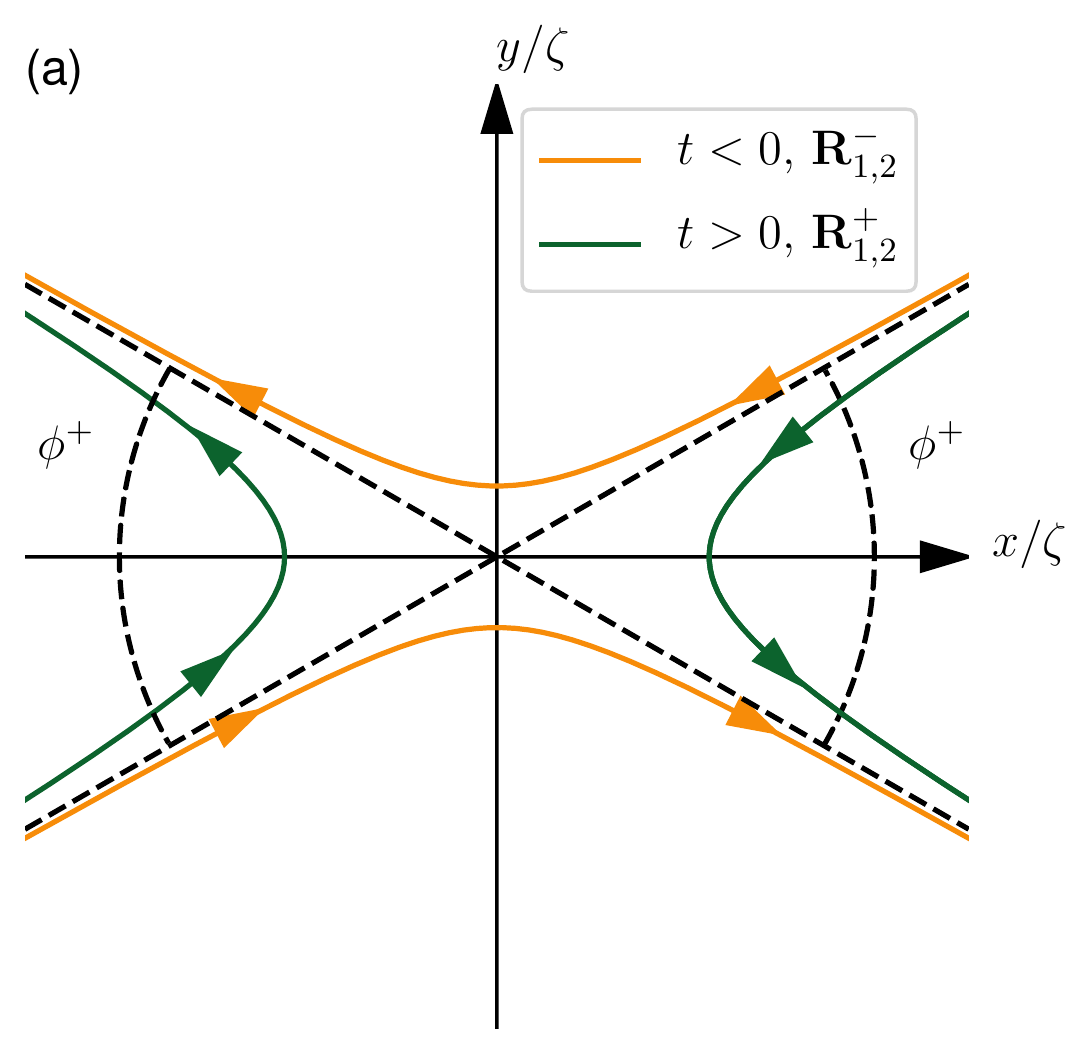}
\quad \quad \quad
\includegraphics[width=0.4\textwidth]{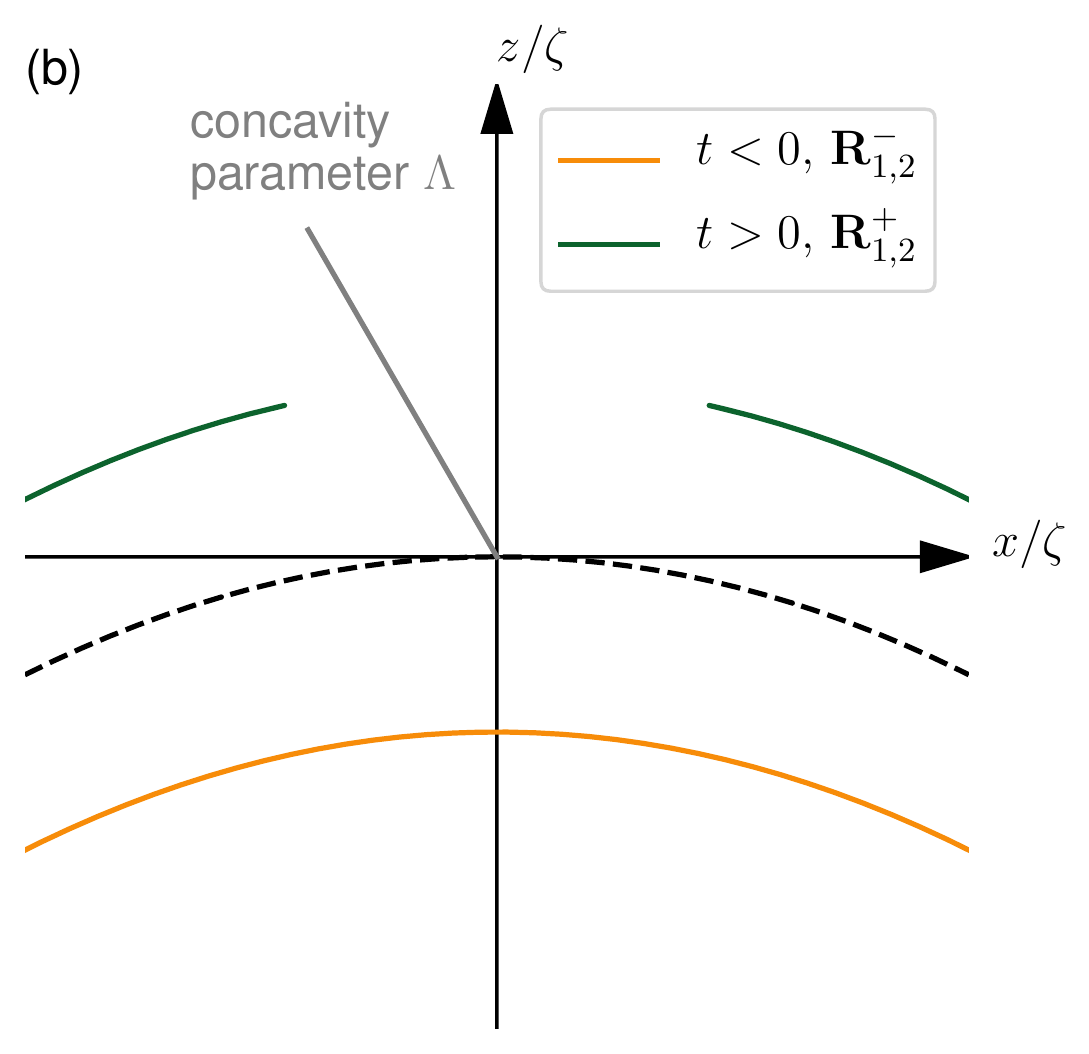}
\caption{(Colour online) 
Sketch of the reconnecting filaments projected \textbf{(a)} onto the $ z=0 $ plane and \textbf{(b)} onto the $ y=0 $ plane.
}
\label{fig:fil}
\end{figure}

We denote by ${\bf R}^-_1$ and ${\bf R}^-_2$ the sets of points of the two vortex filaments before reconnection and by ${\bf R}^+_1$ and ${\bf R}^+_2$ the ones after reconnection. 
Without loss of generality we may assume: that (i) about the reconnection point, ${\bf R}^-_1\subset \{y>0\} $ and ${\bf R}^-_2\subset \{y<0\} $ whereas ${\bf R}^+_1\subset \{x<0\} $ and ${\bf R}^+_2\subset \{x>0\} $; and that (ii) the orientation of the vorticity follows the arrows as in the sketch displayed in Fig.~\ref{fig:fil}(a).
In order to find the range of the admissible values of the wave-function parameters $ p\pm1 $ and $ (A, B, C, D) \in \mathbb{R} $ of eq. (\ref{eq:genIC}), we thus need to impose the following validity conditions.
\begin{itemize}
\item 
{\it Existence of the hyperbola}.
At the reconnection time $ t_{\rm r}=0 $ we want the hyperbola asymptotes 
$ y=\pm\sqrt{(C-A)/(B-D)} \, x $
set by Eq. (\ref{eq:genICProj-xy}) to be real (in other words we want the equation to describe a hyperbola and not an ellipse).
This reduces to the condition
 \begin{equation}
\frac{C-A}{B-D} \ge 0 \, .
\label{eq:condHyp}
\end{equation}
\item 
{\it Convention on the location of the filaments}.
Our convention adopts that, about the reconnection point, the positions of the filaments satisfy ${\bf R}^-_1\subset \{y>0\} $ and ${\bf R}^-_2\subset \{y<0\} $ whereas ${\bf R}^+_1\subset \{x<0\} $ and ${\bf R}^+_2\subset \{x>0\} $.
Hence, by evaluating eq. (\ref{eq:genICProj-xy}) at times $ t<0 $ and $ t>0 $ we obtain that the following conditions, respectively, must hold 
\begin{equation}
(B-D) \, p \,  (C+D+A+B) < 0
\quad \mbox{and} \quad
(C-A) \, p \,  (C+D+A+B) < 0 \, .
\label{eq:condPos}
\end{equation}
\item 
{\it Convention on the vorticity orientation of the filaments}.
The orientation of the filaments can be evaluated by computing the pseudo-vorticity ${\bf \omega}=\nabla Re(\psi)\times \nabla Im(\psi)$ of the wave-function in Eq.~(\ref{eq:solPsi}) at its nodal lines \cite{VilloisTrackingAlgo}.
In order to impose the vorticity orientations as the arrows sketched in Fig.~\ref{fig:fil}(a) the following conditions
\begin{equation}
\left\{
\begin{split}
& p = -1 \\
& B < D \\ 
& A > C \\
\end{split}
\right. 
\quad \mbox{or} \quad
\left\{
\begin{split}
& p = 1 \\
& B > D \\
& A < C \\
\end{split}
\right.    
\label{eq:condVort}
\end{equation}
must be realised.
\end{itemize}

\subsection{The geometrical parameters of the filaments \label{subsec:ParamGeom}}

As shown in the previous subsection, provided that a suitable choice of the wave-function parameters is taken about the reconnection event, the projection of the filaments onto the $ z=0 $ plane corresponds to the hyperbola represented in Eq.~(\ref{eq:genICProj-xy}).
We can define the {\it macroscopic reconnecting angle} $ \phi^+ $ as the angle formed by the hyperbola asymptotes when considering the filaments after the reconnection, as shown in Fig.~{\ref{fig:fil}(a).
By simple geometrical considerations, and using Eq.~(\ref{eq:genICProj-xy}), we find that this angle is related to the wave-function parameters as
\begin{equation}
\phi^+ = 2 \arctan \sqrt{\frac{C-A}{B-D}} 
\quad \Longleftrightarrow \quad
\tan^2 \left(\frac{\phi^+}{2}\right) = \frac{C-A}{B-D}
 \, .
\end{equation}
Also, about the reconnection event, the projection of the filaments onto the $ y=0 $ plane is the parabola found in Eq.~(\ref{eq:genICProj-xz}).
Its concavity results simply in
\begin{equation}
\frac{BC-AD}{(B-D) \zeta} = \frac{\Lambda}{\zeta} \, , \quad \text{given the dimensionless {\it concavity parameter $ \Lambda=\frac{BC-AD}{(B-D)} $. }}
\end{equation}
A sketch of the projection displaying the role of the concavity parameter is shown in Fig.~{\ref{fig:fil}(b).

We can express these two geometrical parameters versus the wave-function parameters characterizing the wave-function initial condition.
A possible choice is the following
\begin{equation}
\left\{
\begin{split}
& A=-B  \tan^2 \left(\frac{\phi^+}{2}\right) + \Lambda \\
& C= -D  \tan^2 \left(\frac{\phi^+}{2}\right) + \Lambda
\end{split}
\right. \, .
\end{equation}
In terms of the geometrical parameters, the region of validity given by Eqs. (\ref{eq:condTheta}), (\ref{eq:condHyp}), (\ref{eq:condPos}), and (\ref{eq:condVort}) results in
\begin{equation}
\left\{  2 \Lambda< \left[\tan^2 \left(\frac{\phi^+}{2}\right)-1\right] (B+D) \right\} \,
\, \cap \, 
 \biggl[
\left( p = -1 \, \cap \, D > B \right) 
\, \cup
\left( p = 1 \, \cap \, D < B \right) \biggr] \, .
\label{eq:condGeom}
\end{equation}
This set of validity conditions are better summarized using the diagrams in Fig.~\ref{fig:condGeom}, where we have set 
\begin{equation} 
T = \tan^2\left(\frac{\phi^+}{2}\right)-1 
\end{equation} 
to simplify the notation.
\begin{figure}
\includegraphics[width=0.4\textwidth]{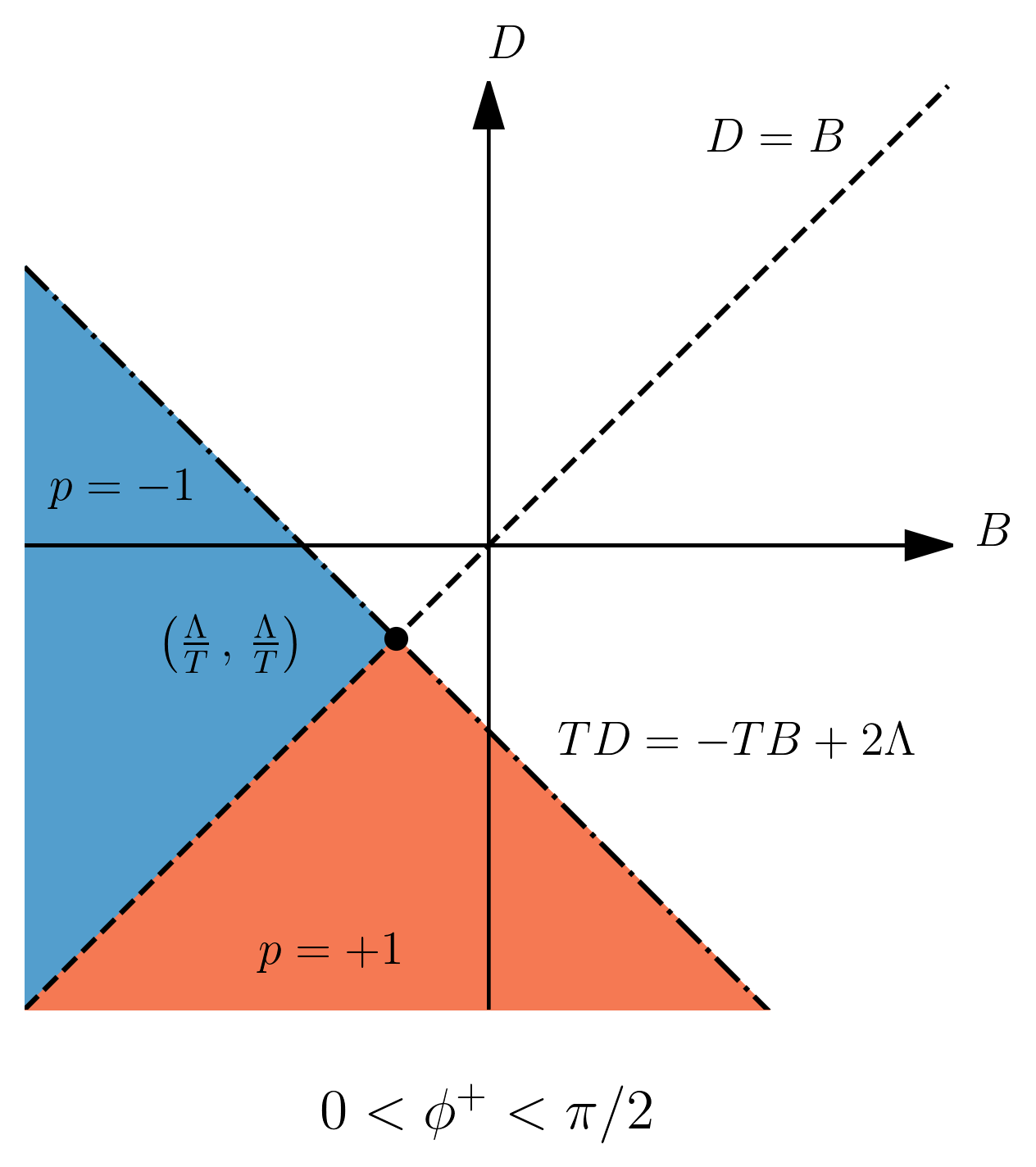}
\quad \quad \quad
\includegraphics[width=0.4\textwidth]{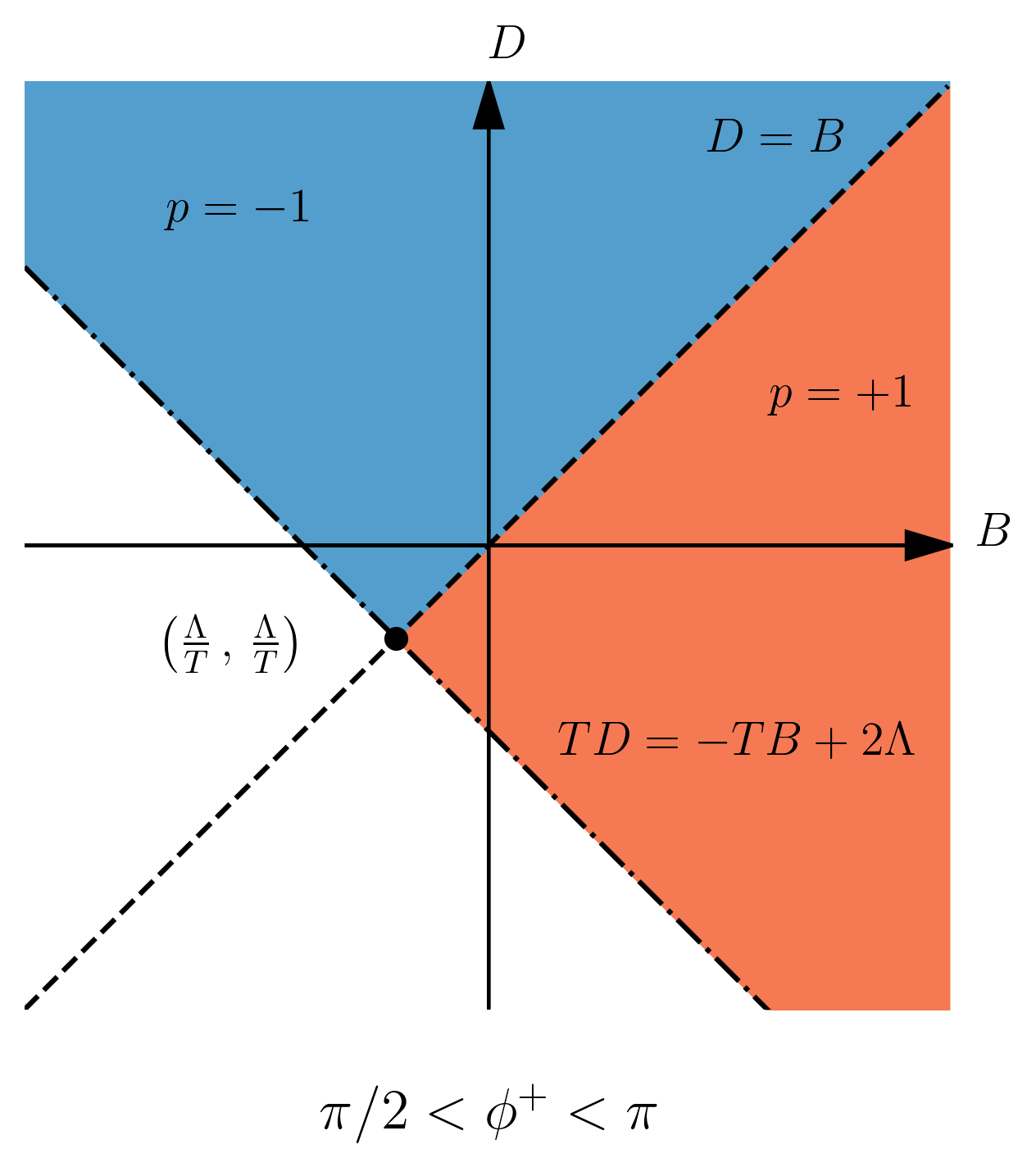}
\caption{(Colour online) Region of the admissible parameters $ (B, D) $ for $ p=-1 $ (blue areas) and $ p=1 $ (red areas), given a specific choice of the geometrical parameters $ (\phi^+, \Lambda) $.}
\label{fig:condGeom}
\end{figure}
For a given choice of the geometrical parameters $ \phi^+ \in (0, \pi/2) \cup (\pi/2, \pi)  $ and $ \Lambda \in \mathbb{R} $, two important remarks should be made: firstly, it is always possible to find a set of values $ (p, B, D) $ that satisfies the validity conditions in Eq.~(\ref{eq:condGeom}); secondly, the area spanned by the parameters $ (B, D) $ is unbounded.
The limit $ \phi^+ \to \pi/2 $, causing $ T\to0 $, needs a special consideration: in this case, using Eq.~(\ref{eq:condGeom}), we find that $ \Lambda \in \mathbb{R}^- $, that is the concavity can only be negative, otherwise the convention on the positions of the filaments is not satisfied. Actually, a completely symmetrical reconnection ($\phi^+=\pi/2$) can not be realized in a fully planar configuration ($\Lambda=0$) with a quadratic wave function.

\subsection{Parametrization of the filaments \label{subsec:paramFil}}
The vortex filaments can be parametrized in terms of the reconnecting angle $\phi^+$ and the concavity parameter $ \Lambda $ as follows:
\begin{eqnarray}
{\bf R}^-_1 (\ell, t) &=&\left\{- \frac{\delta^-(t)}{2} \cot\left(\frac{\phi^+}{2}\right)\sinh{(\ell)},  \frac{\delta^-(t)}{2} \cosh{(\ell)}  ,  z^-(\ell,t) \right\}
\label{eq:filParam1M} \\
{\bf R}^-_2 (\ell, t) &=&\left\{ \frac{\delta^-(t)}{2} \cot\left(\frac{\phi^+}{2}\right)\sinh{(\ell)},  -\frac{\delta^-(t)}{2} \cosh{(\ell)}  ,  z^-(\ell,t) \right\}
\label{eq:filParam2M} \\
{\bf R}^+_1 (\ell, t) &=&\left\{- \frac{\delta^+(t)}{2} \cosh{(\ell)},  \frac{\delta^+(t)}{2} \tan\left(\frac{\phi^+}{2}\right)\sinh{(\ell)}  ,  z^+(\ell,t) \right\}
\label{eq:filParam1P} \\
{\bf R}^+_2 (\ell, t) &=&\left\{ \frac{\delta^+(t)}{2} \cosh{(\ell)},  -\frac{\delta^+(t)}{2} \tan\left(\frac{\phi^+}{2}\right)\sinh{(\ell)}  ,  z^+(\ell,t) \right\},
\label{eq:filParam2P}
\end{eqnarray}
where $ \ell \in (-\infty, +\infty) $ is the parameter spanning the entire length of the filaments (note however that $ \ell $ does not correspond to the arc-length parametrization of the filaments).
Here
\begin{equation}
\delta^\pm(t)=A^\pm\sqrt{\Gamma |t|} \, , \quad 
A^-=\sqrt{\frac{ 2 \left[ \tan^2(\phi^+/2)-1\right] (B+D) - 4 \Lambda}{(B-D) p \pi}} \, , \quad
A^+=\frac{A^-}{ \tan(\phi^+/2)} \, ,
\label{eq:As}
\end{equation}
and $ z^\pm(\ell,t) $ can be expressed by using eq.~\eqref{eq:genICProj-xz} and the square of the $ x $-component of the filaments before and after the reconnection, respectively, resulting in
\begin{equation}
\begin{split}
& z^-(\ell, t)=z_{\rm r}(t) + \frac{\Lambda}{8 \zeta}\left[ \delta^-(t) \cot\left(\frac{\phi^+}{2}\right) \sinh(\ell) \right]^2 \\
& z^+(\ell, t)=z_{\rm r}(t) + \frac{\Lambda}{8 \zeta}\left[ \delta^+(t) \cosh(\ell) \right]^2 \\
\end{split} 
\, , 
\end{equation}
where
\begin{equation}
z_{\rm r}(t) = \frac{-(B^2+D^2)\left[\tan^2(\phi^+/2)-1\right] + (B+D) \Lambda}{4(B-D) p \pi \zeta} \, \Gamma t \, .
\label{eq:zr}
\end{equation}

This choice of parametrization is particularly useful as one can see easily that the distance between the filaments before and after the reconnection is given by
\begin{equation}
|{\bf R}^\pm_1 (\ell=0, t) -{\bf R}^\pm_2 (\ell=0, t) |=\delta^\pm(t) \, ,
\end{equation}
which immediately demonstrates the scaling $ \delta \propto |t-t_{\rm r}|^{1/2} $ \cite{nazarenko2003analytical, villoisPRF2018}. 
Furthermore, we can notice that the ratio between the rate of approach and separation is independent on the concavity parameter $ \Lambda $ and reads
 \begin{equation}
A_{\rm r} = \frac{A^+}{A^-} = \cot\left(\frac{\phi^+}{2}\right)  \, .
\end{equation}
Finally, we can understand by looking at Eqs.~(\ref{eq:As}) and (\ref{eq:zr}) that the choice of the parameters $ (B, D) $ only influences the time-scale of the reconnection process.

\section{Momentum and energy transfers during a reconnection \label{sec:Deltas}}

As it has been observed in previous works \cite{PhysRevLett.86.1410, KerrPRL2011, Zuccher&Caliari&Baggaley&BarenghiPof2012} and clearly displayed in Fig.\ref{fig:densityEvol}, when a vortex reconnection takes place in a quantum fluid, a sound pulse is excited. Energy and momentum are thus transferred from the incompressible to the compressible degrees of freedom of the superfluid in an irreversible manner. The aim of this section, and the main result of this work, is to develop an asymptotic matching theory that allows for quantifying such energy and momentum exchanges.

In the GP model, the total energy \eqref{eq:H} and linear momentum \eqref{eq:P-GP} are conserved during the reconnection process, as they are integrals of motion. Well before a reconnection event, practically only the presence of vortex filaments contribute to the invariants, whereas after reconnection, both filaments and compressible waves add up their contributions to them. If one is able to estimate the contribution of the filaments, then the contribution of density waves can be deduced using the conservation of the invariants. In the case of energy, such decomposition can be easily done numerically by splitting the kinetic energy term into the incompressible and compressible parts \cite{nore1997decaying}. Such measurements were performed in \cite{Villois2020Irreversible} and will be reproduced below in our discussions.

Our analytical treatment of the problem is as follows. When the filaments are far from each other, i.e. $\delta(t)^\pm\gg \xi$, their dynamics of mainly driven by the Biot-Savart model. In that region we might use the vortex filament description to evaluate their energy and momentum of the superfluid. On the other hand, when $\delta^\pm(t)\ll\xi$, the dynamics is governed by the linear regime given by the Schroedinger equation. Vortices then reconnect following the laws described in the preceding section. We thus describe the reconnection matching, sketched in Fig.\ref{fig:matching}, as follows. Before reconnection, some Biot-Savart dynamics leads to the pre-reconnection input configuration ${\bf R}^-_{1,2}$ for the filaments about the vortex reconnection point. The Biot-Savart description is assumed to be valid down to a distance $\delta^-=\delta_{\rm lin}$, where $ \delta_{\rm lin} $ is of the order of a few healing lengths. From there, the filaments are driven by the Schroedinger equation allowing them to reconnect. After the reconnection, this linear regime is valid until the vortices separate up to a distance $\delta^+=\delta_{\rm lin}$. The linear evolution thus provides the output post-reconnection configuration ${\bf R}^+_{1,2}$ for the filaments. From there onwards, the dynamics is again governed by the Biot-Savart model. 
\begin{figure}[h!]
\includegraphics[width=0.8\textwidth]{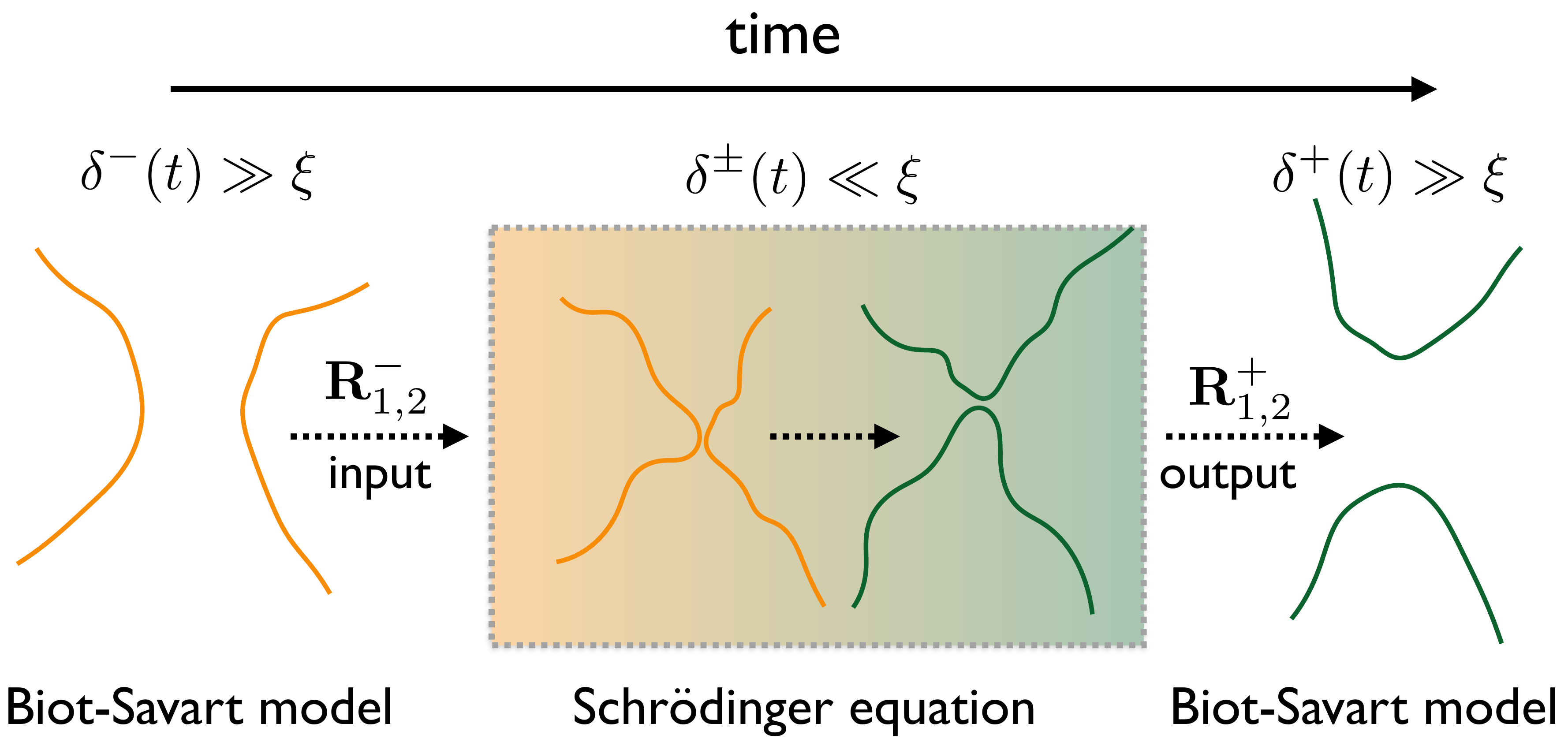} 
\caption{(Colour online) 
A sketch of a reconnection process and matching asymptotics. When vortices are far apart their dynamics if governed by the Biot-Savart equation, whereas when they are about to reconnect the process is driven by the Schroedinger equation.
}
\label{fig:matching}
\end{figure}
Note that the linear regime corresponds only to the dynamics inside the orange-greenish box in Fig.~\ref{fig:matching}. We can thus consider the linear regime as the regularization mechanism allowing vortex reconnections in the Biot-Savart model.

Summarizing, in order to compute the differences before and after the reconnection in the incompressible energy and momentum of the superfluid, we use the theoretical description ${\bf R}^\pm_{1,2}$ for the filaments given in Eqs.~(\ref{eq:filParam1M}-\ref{eq:filParam2P}). Namely, we use such parametrization when the distance is
\begin{equation}
\delta^-=\delta^+=\delta_{\rm lin}\gtrsim\xi \, ,
\end{equation}
as illustrated in Figs.~\ref{fig:recSphere}(a) and \ref{fig:recSphere}(b), respectively.
\begin{figure}]
\includegraphics[width=0.35\textwidth]{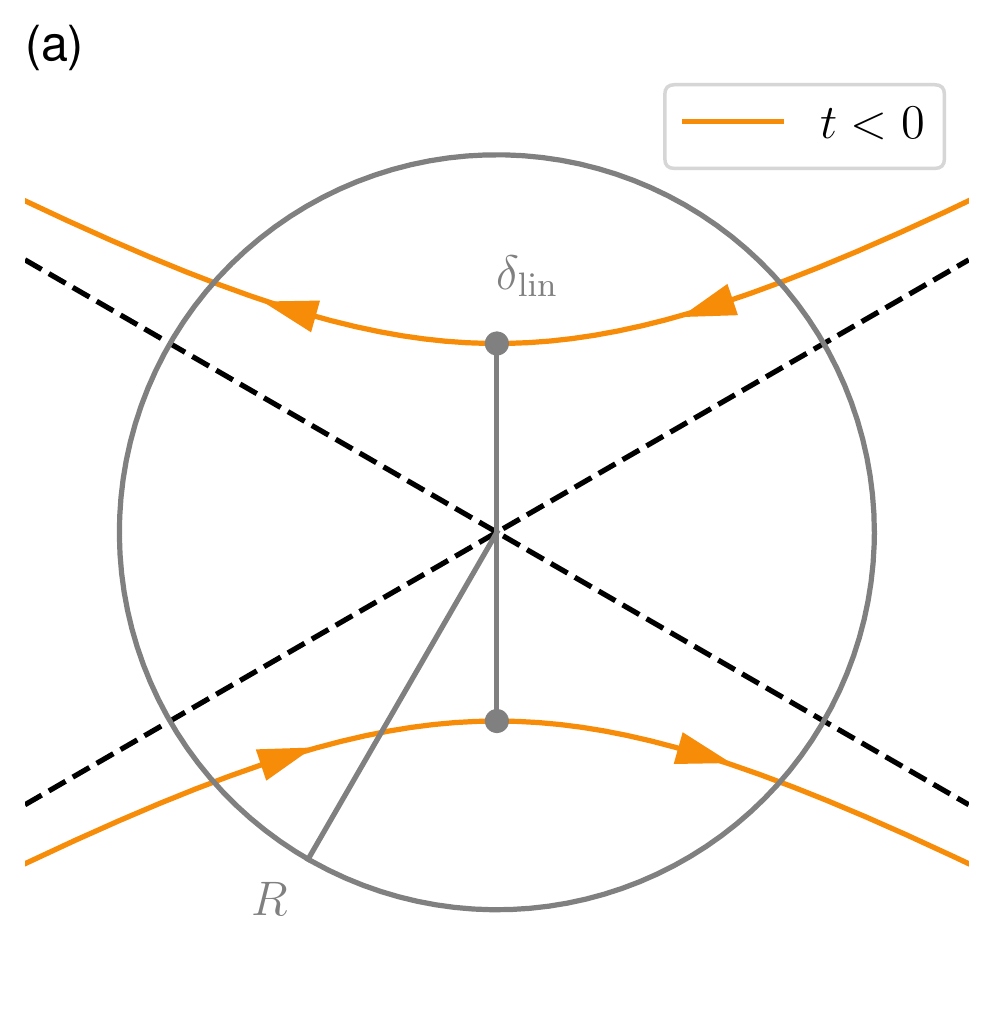} 
\quad \quad \quad 
\includegraphics[width=0.35\textwidth]{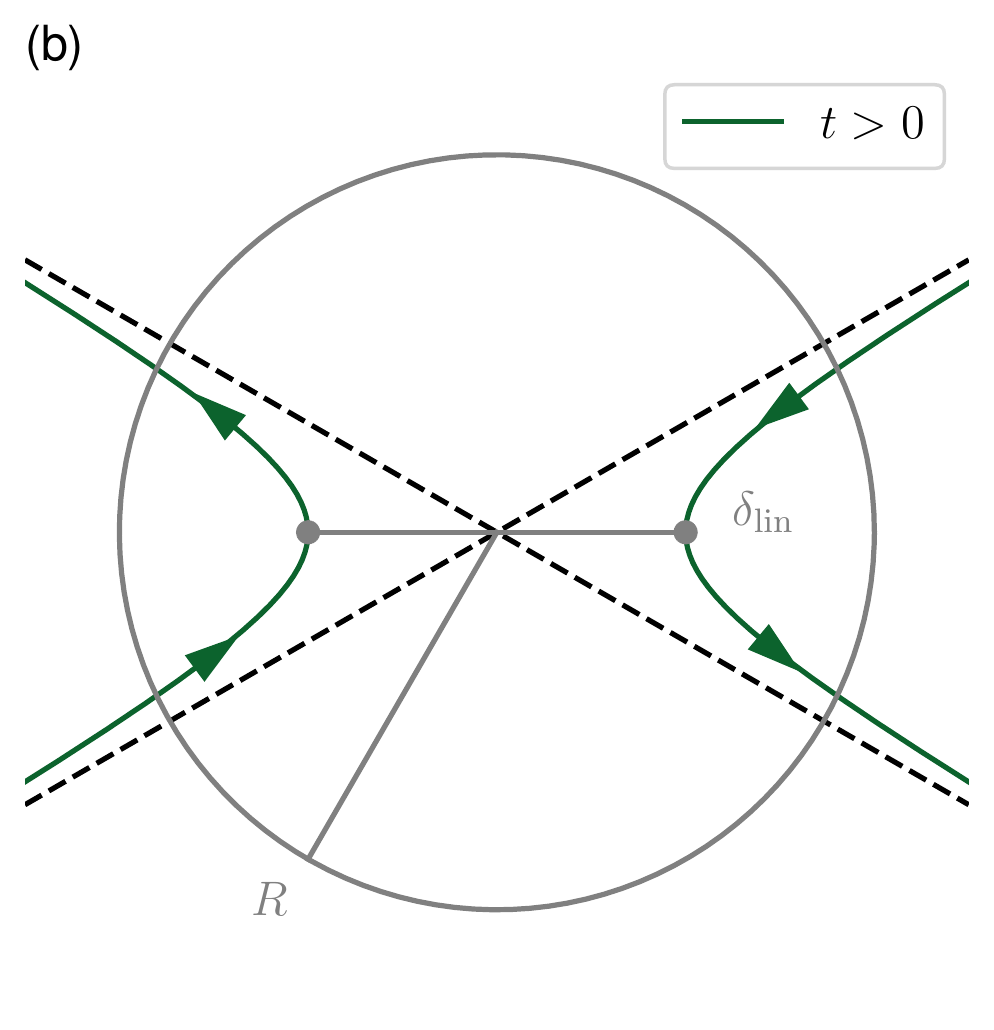} 
\caption{(Colour online) 
Projections of the reconnecting filaments and the cylinder onto the plane $ z=0 $ before the reconnection \textbf{(a)} and after the reconnection \textbf{(b)}.
Here the reconnection angle is $ \phi^+ = \pi/3 $.
}
\label{fig:recSphere}
\end{figure}
Note that the assumption that the linear regime description may be still valid at distances beyond the healing length $ \xi $ is justified by numerical evidence \cite{villoisPRF2018}. 

\subsection{The cylindrical region of integration \label{subsec:cyl}}

As detailed in the following, the calculations of the linear momentum and energy of the vortices involve the integration over the full length of the filaments. As we are interested in their differences, to simplify the problem we will consider only the segments of the filaments which lie inside the cylinder of circle of radius $ R $ centered at the origin and having the cylindrical axis parallel to the $ z $ axis; the projection of the cylinder onto the $ z=0 $ plane, corresponding to the circle of radius $ R $ centered at the origin, is also sketched in Fig.~\ref{fig:recSphere}.

The vortex filaments lie inside the cylinder when their parametrization satisfies $ |\ell|\le L^\pm$ before and after the reconnection, respectively, given
\begin{eqnarray}
L^-(R/\delta_{\rm lin})&=
& \frac{1}{2} \ln\left\{ \frac{8 (R/\delta_{\rm lin})^2 + (A_{\rm r}^2-1) + 2 \sqrt{ \left[ 4 \left(R/\delta_{\rm lin}\right)^2 - 1 \right] \left[4 \left(R/\delta_{\rm lin}\right)^2+A_{\rm r}^2 \right]}}{A_{\rm r}^2+1} \right\}
\\
L^+(R/\delta_{\rm lin}) &=
& \frac{1}{2} \ln\left\{ \frac{ 8 A_{\rm r}^2 (R/\delta_{\rm lin})^2 + (1-A_{\rm r}^2) + 2 A_{\rm r} \sqrt{ \left[ 4 \left(R/\delta_{\rm lin}\right)^2-1 \right] \left[ 4 A_{\rm r}^2 \left(R/\delta_{\rm lin}\right)^2 + 1 \right]} }{A_{\rm r}^2+1} \right\} \, .
\end{eqnarray}
Physically, it is natural to assume that $ R $ is larger, or much larger, than $ \delta_{\rm lin} $. 
Note that, keeping $ A_{\rm r} $ finite, we have
\begin{equation}
\begin{split}
& \lim_{R/\delta_{\rm lin} \to\infty} L^-(R/\delta_{\rm lin}) = \ln(R/\delta_{\rm lin}) + \ln\left( \frac{4}{\sqrt{1+A_{\rm r}^2}}\right) +\ldots \\
& \lim_{R/\delta_{\rm lin} \to\infty} L^+(R/\delta_{\rm lin}) = \ln(R/\delta_{\rm lin}) + \ln\left( \frac{4}{\sqrt{1+A_{\rm r}^2}}\right) + \ln(A_{\rm r}) +\ldots 
\end{split}
\label{eq:LRinf}
\end{equation}

\subsection{Linear momentum difference \label{subsec:deltaP}}

Following L.~Pismen \cite{Pismen:1999aa}, the linear momentum of an incompressible and inviscid fluid with filamentary vorticity field of intensity $ \Gamma $, that is within the framework of the Biot-Savart model, reads
\begin{equation}
\mathbf{P}_{\rm fil} = \frac{\Gamma}{2} \oint \mathbf{R} \times d\mathbf{R} = \int_{\mathcal{L}} \mathbf{p}_{\rm fil}(\ell) \,  d\ell \, ,
  \quad \text{where} \quad \mathbf{p}_{\rm fil}(\ell)= \frac{\Gamma}{2} \sum_\text{filaments} \mathbf{R} \times \frac{\partial \mathbf{R}}{\partial \ell} 
\end{equation}
is the momentum density (per unit of filament length) and the integration interval follows the parametrization of the filaments.
In our case $ \ell \in (-\infty, +\infty) $ and there are only two filaments before and two filaments after the reconnection.
The incompressible superfluid momenta density before and after reconnection can be therefore computed by using Eqs.~(\ref{eq:filParam1M}-\ref{eq:filParam2P}).
Due to the particular symmetries of the reconnecting configuration, their expressions are rather simple.
Interestingly they are independent of $ \Lambda $ and $ \ell $, and read
\begin{equation}
\begin{split}
& \mathbf{p}_{\rm fil}^-  = \left( 0, \, 0, \, \frac{{\delta^-}^2 A_{\rm r} \Gamma}{4}  \right) 
\quad \Longrightarrow \quad 
\mathbf{P}_{\rm fil}^-(\delta^-, A_{\rm r}, R/\delta_{\rm lin})= \left( 0, \, 0, \, \frac{{\delta^-}^2 A_{\rm r} \Gamma}{4}  \right) \int_{-L^-(R/\delta_{\rm lin})}^{L^-(R/\delta_{\rm lin})} d\ell =  \left[ 0, \, 0, \, \frac{{\delta^-}^2 A_{\rm r} \Gamma}{2} L^-(R/\delta_{\rm lin})  \right]  \\
& \mathbf{p}_{\rm fil}^+ = \left( 0, \, 0, \, -\frac{{\delta^+}^2 \Gamma}{4 A_{\rm r}}  \right) 
\quad \Longrightarrow \quad 
\mathbf{P}_{\rm fil}^+(\delta^+, A_{\rm r}, R/\delta_+)= \left( 0, \, 0, \, -\frac{{\delta^+}^2 \Gamma}{4 A_{\rm r}}  \right) \int_{-L^+(R/\delta_{\rm lin})}^{L^+(R/\delta_{\rm lin})} d\ell =  \left[ 0, \, 0, \, -\frac{{\delta^+}^2 \Gamma}{2 A_{\rm r}} L^+(R/\delta_{\rm lin}) \right] \, . 
\end{split}
\end{equation}
These results immediately tell us that the difference of the linear momentum of vortex filaments during a reconnection is non zero only along the $z$-direction and in the limit of large $ R/\delta_{\rm lin}$ is given by 
\begin{equation}
\Delta P_{{\rm fil}, z}(\delta_{\rm lin}, A_{\rm r}, R/\delta_{\rm lin}) 
 = P_{{\rm fil}, z}^+(\delta_{\rm lin}, A_{\rm r}, R/\delta_{\rm lin}) - P_{{\rm fil}, z}^-(\delta_{\rm lin}, A_{\rm r}, R/\delta_{\rm lin})
 \propto -\frac{1+A_{\rm r}^2}{A_{\rm r}} = -2\csc\phi^+ \, .
 \label{eq:DeltaPz}
\end{equation}

This is a remarkable result: following our convention, refer again to Fig.~\ref{fig:fil} and for all $ (\delta_{\rm lin}, A_{\rm r}, R/\delta_{\rm lin}) $, the incompressible superfluid linear momentum changes during a reconnection only along the $ z $-axis and, most importantly, its variation is always negative with a global maximum for $ \phi^+=\pi/2 $.
As a consequence, assuming that the total linear momentum within the cylinder of radius $ R $ remains conserved during the reconnection process, a density/phase compressible excitation, a sound pulse, must carry the missing momentum
\begin{equation} 
\mathbf{P}_{\rm pulse} = -\Delta \mathbf{P}_{\rm fil} \, ,
\end{equation}
that is, must certainly move along the positive direction of the $ z $-axis.

\subsection{Energy difference \label{subsec:deltaE}}
In first approximation, the (kinetic) energy of an incompressible inviscid fluid consisting on a collection of vortex filaments is proportional to the total length $ \mathcal{L} $ of the vortex configuration.
This result arises from the local induction approximation (LIA) of the Biot-Savart model. The (kinetic) incompressible energy of the superfluid, in the LIA assumption, is thus given by
\begin{equation}
E_{\rm LIA}(\mathcal{L}) = \mathcal{T} \mathcal{L}= \mathcal{T}\int_\mathcal{L} \sum_{\text{filaments}} \left|\frac{\partial {\bf R}}{\partial \ell}\right| \, d\ell \, ,
\end{equation}
where $\mathcal{T}=\rho_0\Gamma^2 \log(L_0/a_0)/(4 \pi^2)$ is the so-called vortex line tension \cite{soninVortexOscillationsHydrodynamics1987,Pismen:1999aa}, with $ L_0 $ is a characteristic length order of the mean radius and $ a_0 $ is the vortex core size. The precise definition of the vortex tension is not relevant for the next considerations.

The incompressible superfluid energy, before and after the reconnection, thus results in
 \begin{equation}
E_{\rm LIA}^\pm(A_{\rm r}, \Lambda/\zeta, \delta_{\rm lin}, R/\delta_{\rm lin}) =\mathcal{T} \int_{-L^\pm(R/\delta_{\rm lin})}^{L^\pm(R/\delta_{\rm lin})} \left|\frac{\partial {\bf R}_1^\pm}{\partial \ell}\right|+ \left|\frac{\partial {\bf R}_2^\pm}{\partial \ell}\right| \, d\ell  \, ,
\end{equation}
and the energy difference during the reconnection is simply
\begin{equation}
\Delta E(A_{\rm r}, \Lambda/\zeta, \delta_{\rm lin}, R/\delta_{\rm lin}) = E^+ (A_{\rm r}, \Lambda/\zeta, \delta_{\rm lin}, R/\delta_{\rm lin}) - E^-(A_{\rm r}, \Lambda/\zeta, \delta_{\rm lin}, R/\delta_{\rm lin}) \propto \Delta\mathcal{L}(A_{\rm r}, \Lambda/\zeta, \delta_{\rm lin}, R/\delta_{\rm lin}) \, ,
\label{eq:enDiff}
\end{equation}
where
\begin{equation} 
\Delta\mathcal{L} (A_{\rm r}, \Lambda/\zeta, \delta_{\rm lin}, R/\delta_{\rm lin}) =  \int_{-L^+(R/\delta_{\rm lin})}^{L^+(R/\delta_{\rm lin})} \left|\frac{\partial {\bf R}_1^+}{\partial \ell}\right|+ \left|\frac{\partial {\bf R}_2^+}{\partial \ell}\right| \, d\ell  - \int_{-L^-(R/\delta_{\rm lin})}^{L^-(R/\delta_{\rm lin})} \left|\frac{\partial {\bf R}_1^-}{\partial \ell}\right|+ \left|\frac{\partial {\bf R}_2^-}{\partial \ell}\right| \, d\ell 
\label{eq:lenDiff}
\end{equation}
is the difference of the total length of the filaments during the reconnection process.

The integrals in (\ref{eq:lenDiff}) can be evaluated analytically for $ \Lambda=0 $ and the final result reads
\begin{equation}
\Delta \mathcal{L}(A_{\rm r}, \Lambda/\zeta=0, \delta_{\rm lin}, R/\delta_{\rm lin}) = 
\frac{2 i \delta_{\rm lin}}{A_{\rm r}} 
\left\{ A_{\rm r}^2 {\rm Ell} \left[ i L^+(R/\delta_{\rm lin}) \left|1+\frac{1}{A_{\rm r}^2}\right.\right]
- {\rm Ell} \left[ i L^-(R/\delta_{\rm lin}) |A_{\rm r}^2+1 \right] \right\} \, ,
\label{eq:enDiffLambda0}
\end{equation}
where $ {\rm Ell}(\cdot | \cdot) $ is the incomplete elliptic integral of the second kind.
The behavior of eq.~(\ref{eq:enDiffLambda0}) for different values of $ R/\delta_{\rm lin} $ is shown in Fig.~\ref{fig:deltaL0}(a).
\begin{figure}
\raggedright (a) \hspace{250pt} (b)  \\
\centering
\includegraphics[width=0.45\textwidth]{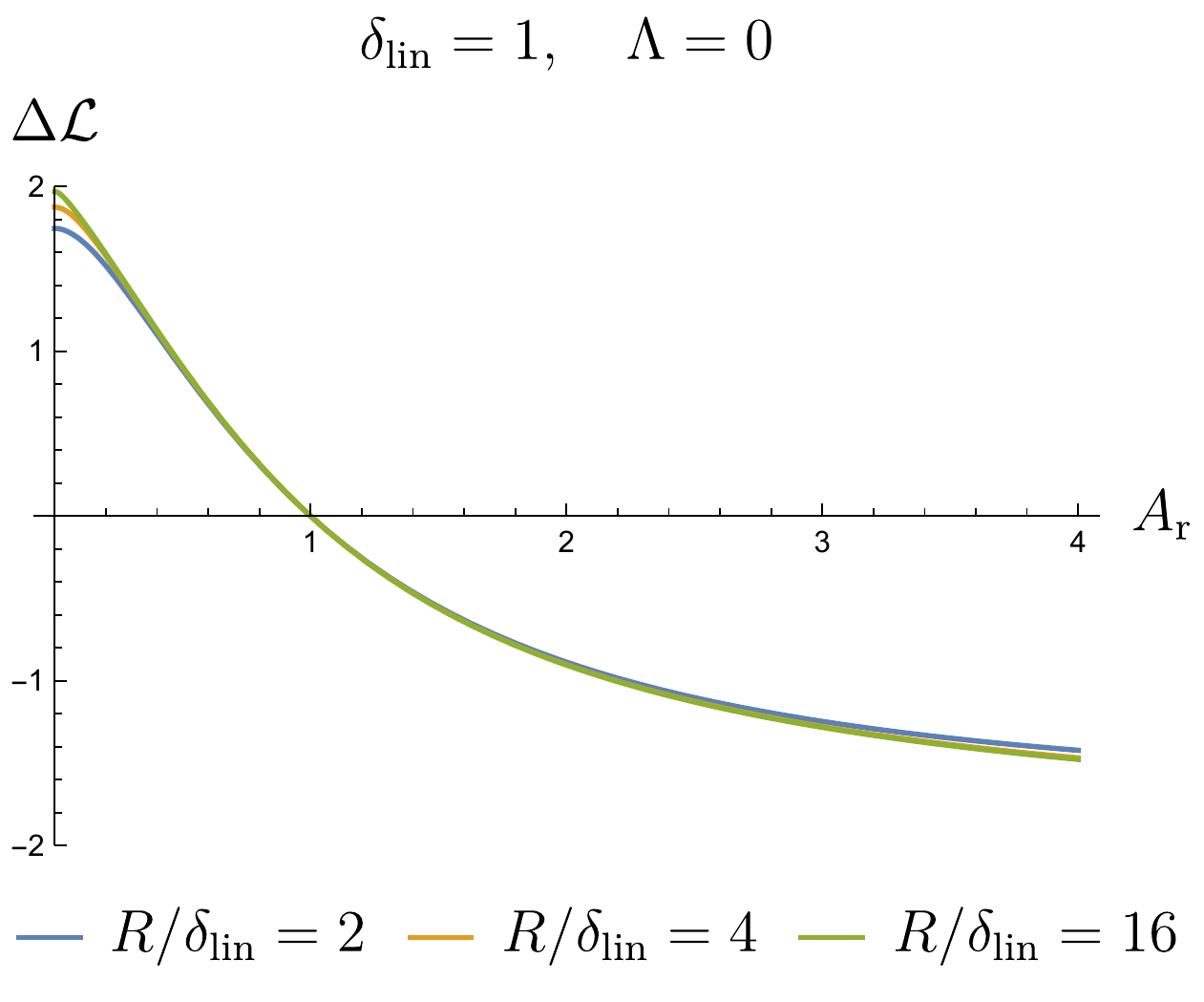}
\quad
\includegraphics[width=0.45\textwidth]{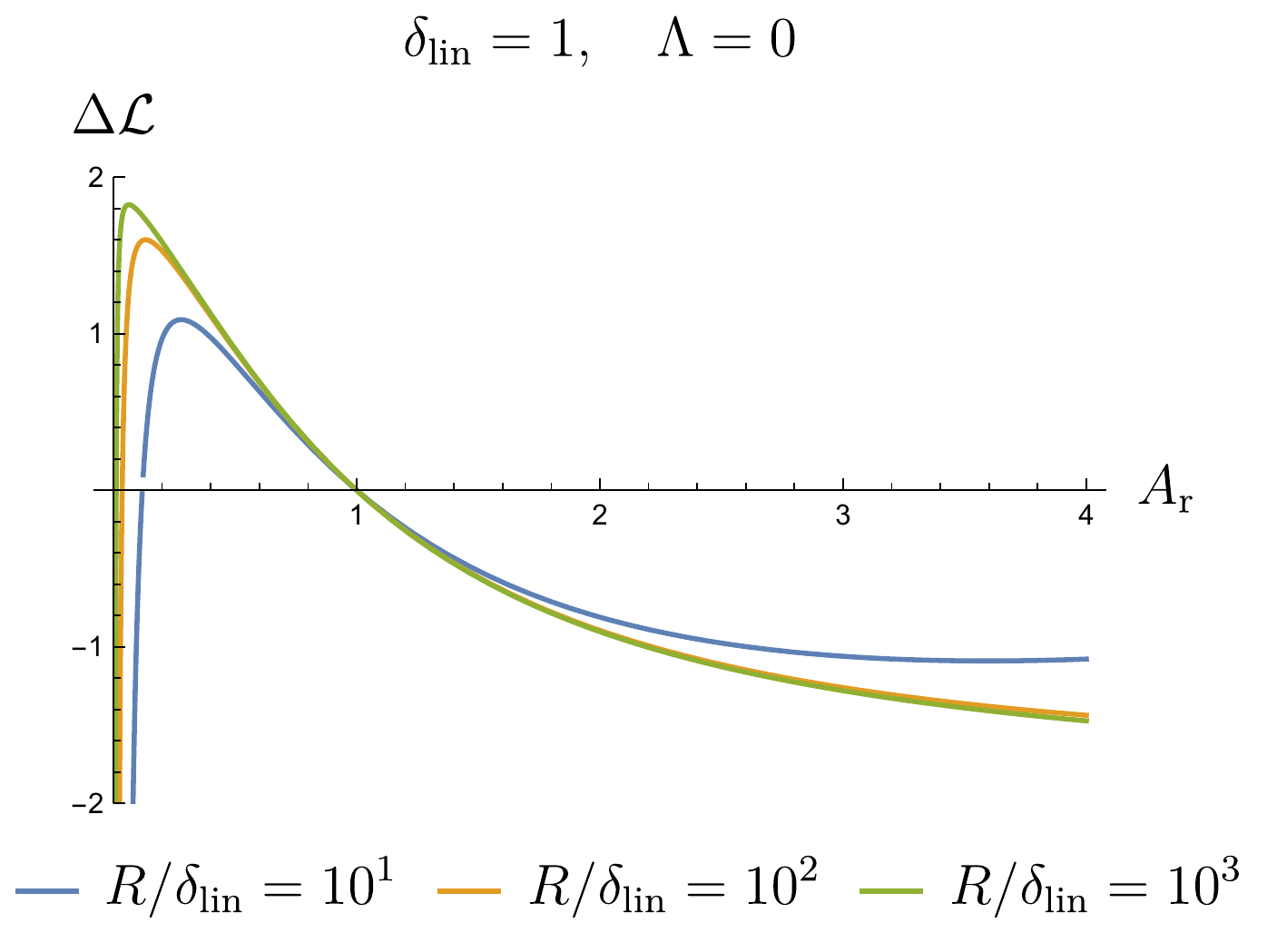}
\caption{(Colour online)
Plots of the difference in the total length of the filaments during the reconnection process when $ \Lambda=0 $ and its limiting case when $ R/\delta_{\rm lin} \gg 1 $, that is eq.s~(\ref{eq:enDiffLambda0}) and (\ref{eq:enDiffLambda0Rinf}), versus $ A_{\rm r} $ for different values of $ R/\delta_{\rm lin} $; here $ \delta_{\rm lin} = 1 $.}
\label{fig:deltaL0}
\end{figure}
Note that $ \Delta\mathcal{L} $ appears to be monotonically decreasing versus $ A_{\rm r} $ and to saturate for large $ R/\delta_{\rm lin} $.
This latter property can be corroborated by taking the limit $ R/\delta_{\rm lin} \to \infty $ keeping $ A_{\rm r} $ finite: using eq.~(\ref{eq:LRinf}) one gets
\begin{equation}
\Delta \mathcal{L}(A_{\rm r},\Lambda/\zeta = 0, R/\delta_{\rm lin} \to\infty) = 
\frac{2 i \delta_{\rm lin}}{A_{\rm r}} 
\left( A_{\rm r}^2 {\rm Ell} \left\{ i \ln \left[\frac{4 \left(R/\delta_{\rm lin}\right)}{\sqrt{1+A_{\rm r}^2}} \right] \left|1+\frac{1}{A_{\rm r}^2}\right.\right\}
- {\rm Ell} \left\{ i \ln \left[\frac{4 A_{\rm r} \left(R/\delta_{\rm lin}\right)}{\sqrt{1+A_{\rm r}^2}} \right] |A_{\rm r}^2+1 \right\} \right) \, .
\label{eq:enDiffLambda0Rinf}
\end{equation}
Figure~\ref{fig:deltaL0}(b) shows its behavior versus $ A_{\rm r} $ for larger and larger values of $ R/\delta_{\rm lin} $; one has however to be cautious in using this limiting expression, as Eq.~(\ref{eq:enDiffLambda0Rinf}) presents an unphysical behavior when $ A_{\rm r} \to 0 $ which is absent in the original Eq.~(\ref{eq:enDiffLambda0}).

Also, we can note that due to symmetry reasons, the two following properties hold 
\begin{equation}
\Delta \mathcal{L}(A_{\rm r}=1,\Lambda/\zeta, \delta_{\rm lin}, R/\delta_{\rm lin})=0 
\label{eq:enDiffAr1}
 \end{equation}
 and
 \begin{equation}
 \Delta \mathcal{L}(A_{\rm r}, \Lambda/\zeta, \delta_{\rm lin}, R/\delta_{\rm lin}) = \Delta \mathcal{L}(A_{\rm r}, -\Lambda/\zeta, \delta_{\rm lin}, R/\delta_{\rm lin}) \, .
 \end{equation}
 
In the general case where $ \Lambda \neq 0 $ and finite, the integrals in (\ref{eq:lenDiff}) have to be computed numerically.
As for $ \Lambda=0 $, $ \Delta\mathcal{L} $ appears to converge in the limit $ R/\delta_{\rm lin} \to\infty $; also, as shown in Figs.~\ref{fig:deltaL}(a) and \ref{fig:deltaL}(b),
\begin{figure}
\raggedright (a) \hspace{250pt} (b)  \\
\centering
\includegraphics[width=0.45\textwidth]{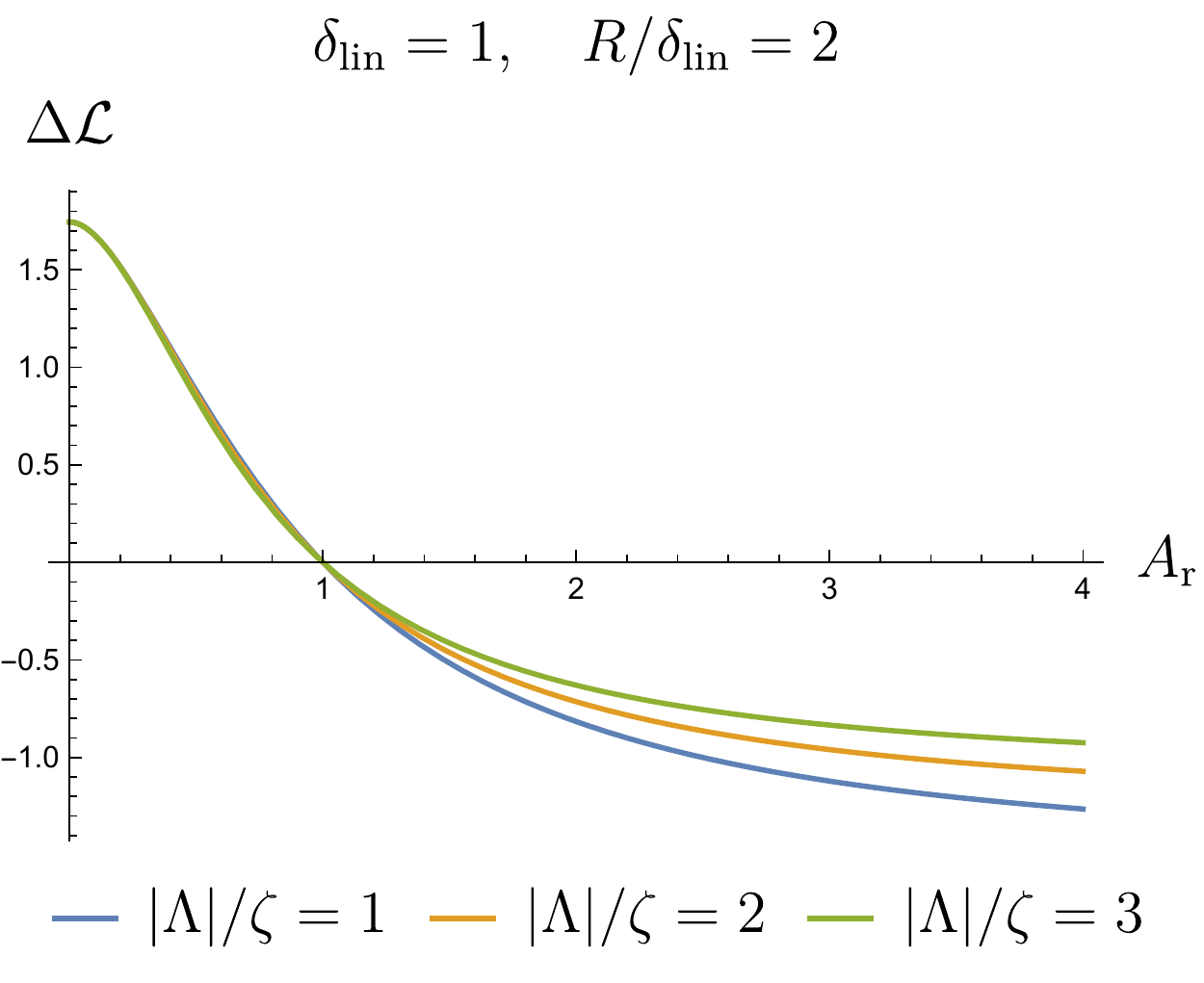}
\quad
\includegraphics[width=0.45\textwidth]{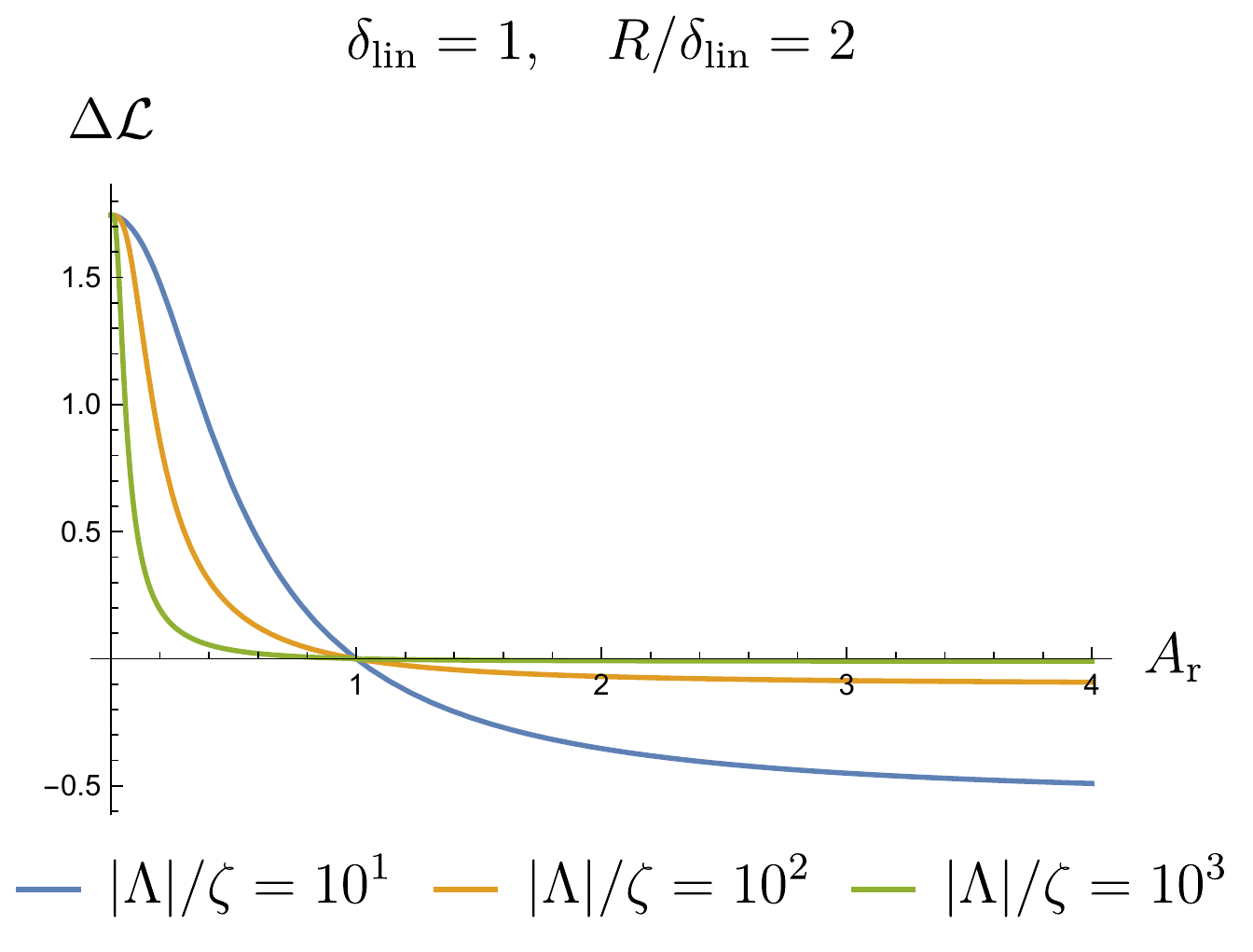}
\caption{(Colour online)
Plots of the difference in the total length of the filaments during the reconnection process, that is a numerical estimation of eq~(\ref{eq:lenDiff}), versus $ A_{\rm r} $ for different values of $ |\Lambda|/\zeta $; here $ \delta_{\rm lin} = 1 $ and $ R/\delta_{\rm lin} =2 $. }
\label{fig:deltaL}
\end{figure}
we observe numerically that
\begin{equation}
\Delta \mathcal{L}(A_{\rm r} \neq 0 , |\Lambda|/\zeta\to\infty, \delta_{\rm lin}, R/\delta_{\rm lin})=0 \, .
\label{eq:enDiffLambdaInf} 
\end{equation}
Hence, let us note that the value of $ \Delta\mathcal{L}(A_{\rm r}, \Lambda, \delta_{\rm lin}, R/\delta_{\rm lin}) $ is always bounded by Eqs.~(\ref{eq:enDiffLambda0}) and (\ref{eq:enDiffLambdaInf}).

Using the same rationale as in the linear momentum difference subsection, we now assume that the total (kinetic) superfluid energy within the cylinder is conserved during the reconnection. 
Hence, to enforce this conservation, a density/phase compressible excitation, a sound pulse, must be created with energy fraction
\begin{equation}
\frac{E_{\rm pulse} }{E_{\rm tot}}= - \frac{\Delta E_{\rm LIA}(A_{\rm r}, |\Lambda/\zeta|, \delta_{\rm lin}, R/\delta_{\rm lin})}{E_{\rm LIA, 0}} = - \frac{\Delta \mathcal{L}(A_{\rm r}, |\Lambda/\zeta|, \delta_{\rm lin}, R/\delta_{\rm lin})}{\mathcal{L}_0} \, ,
\label{eq:Ewav}
\end{equation}
given $ E_{\rm tot} $ the total superfluid energy, $ E_{\rm LIA, 0} $ is the LIA energy of the initial configuration of the filaments, and $ \mathcal{L}_0 $ the total initial length of the filaments.
As $ \Delta\mathcal{L} $ appears to be monotonically decreasing versus $ A_{\rm r} $ and equal to zero at $ A_{\rm r}=1 $, then $ \Delta E_{\rm pulse} $ is positive only for $ A_{\rm r} > 1 $.
This is a very important result, as it shows that, under our assumptions, reconnections cannot happen for $ A_{\rm r} \le 1 $, that is for $ \phi^+ \ge \pi/2 $, because a sound pulse with negative energy should be created, which is clearly unphysical.

\section{Discussions and future perspectives \label{sec:concl} }
Summarizing, following the calculations presented in Section~\ref{sec:Deltas}, we are able to estimate the linear momentum and energy of the sound pulse emitted during a reconnection event.
These reads
\begin{equation}
\begin{split}
& \mathbf{P}_{\rm pulse} = \left[0, 0, -\Delta P_{{\rm fil}, z}(\delta_{\rm lin}, A_{\rm r}, R) \right] \\
& E_{\rm pulse} = -\frac{\Delta\mathcal{L}(A_{\rm r}, |\Lambda|/\xi, \delta_{\rm lin}, R/\delta_{\rm lin})}{\mathcal{L}_0} \, E_{\rm tot}
\end{split} \, ,
\label{eq:DeltasPulseFil}
\end{equation}
where the details of the functions can be found in Eqs (\ref{eq:DeltaPz}) and (\ref{eq:lenDiff}), respectively.
In particular, we underline that for all $ (|\Lambda|/\xi, \delta_{\rm lin}, R/\delta_{\rm lin}) $
\begin{equation}
\begin{split}
& P_{{\rm pulse}, z} > 0 \, , \quad \text{for all $ A_{\rm r} $} \\
& E_{\rm pulse} > 0 \, , \quad \text{for $ A_{\rm r}>1 $}
\end{split} \, ,
\end{equation}
explain the origin of the asymmetry observed in the distribution of the the pre-factors $ A^\pm $, and the directionality observed in the sound pulse emission.

Let us now to discuss further the nature of the sound pulse. As verified in reference \cite{Villois2020Irreversible} and apparent in Fig.\ref{fig:densityEvol}, the pulse indeed propagates in a well defined direction. Figure~\ref{Fig:Pulse}(a) shows such a sound pulse, rescaled by the bulk density, propagating along the positive $ z $-direction after a reconnection characterized by the geometrical parameter $ A_{\rm r}=1.67 $. The pulse is plotted versus time in Fig.\ref{Fig:Pulse}.a and versus the retarded time in Fig.\ref{Fig:Pulse}.b for different values of the $z$-coordinate.
\begin{figure}
\includegraphics[width=0.3\textwidth]{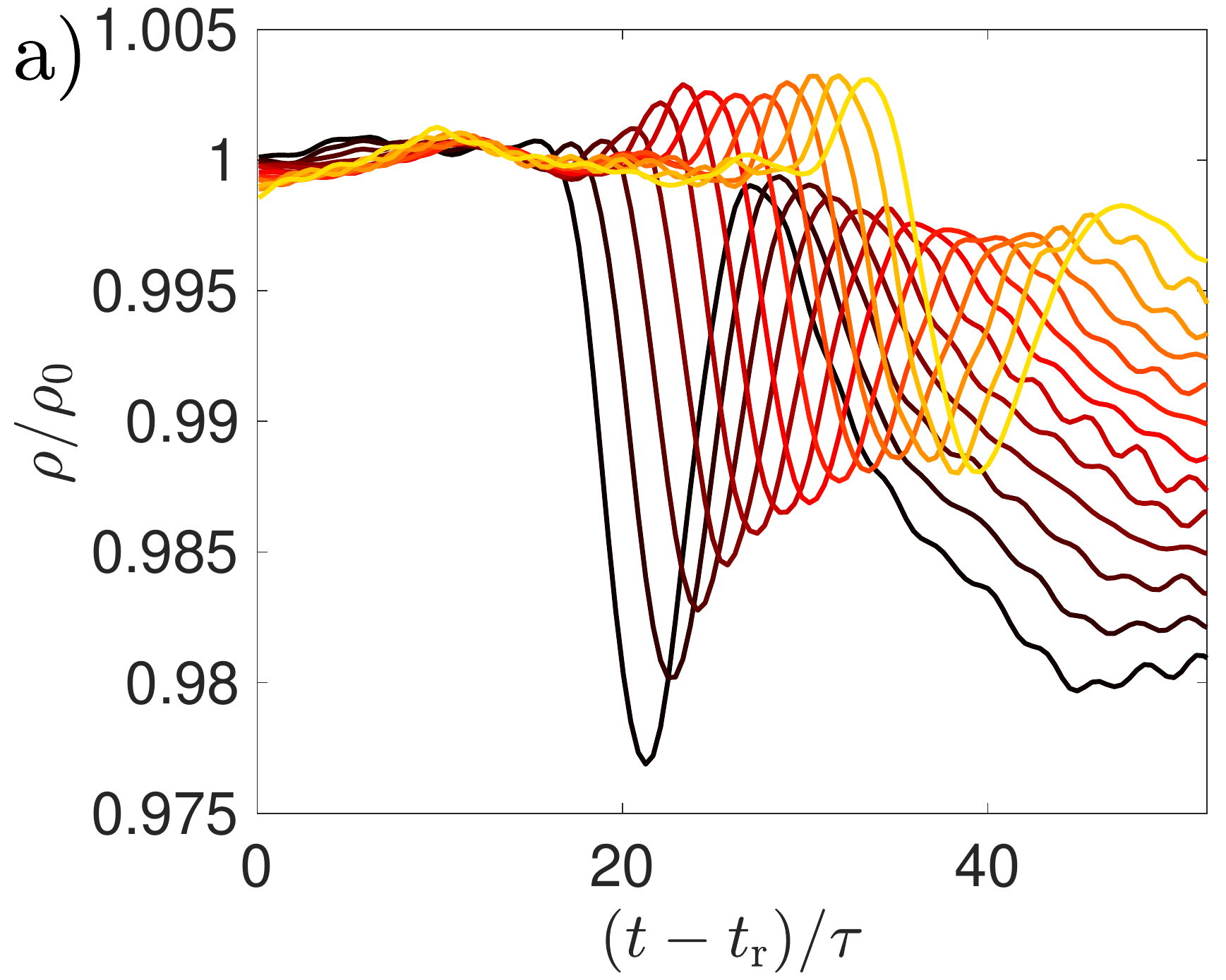} 
\includegraphics[width=0.079\textwidth]{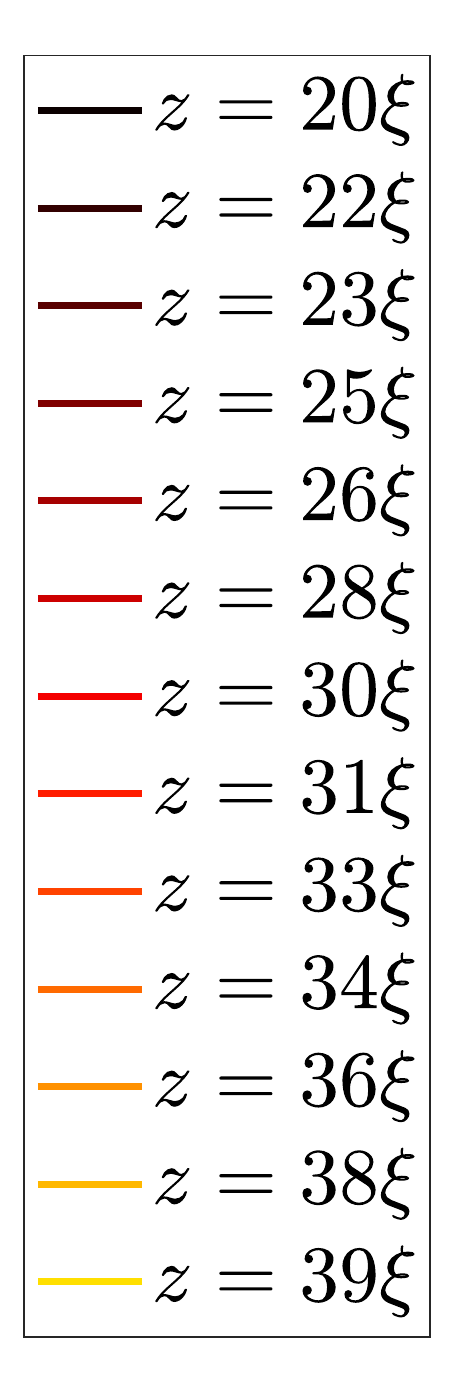} 
\includegraphics[width=0.3\textwidth]{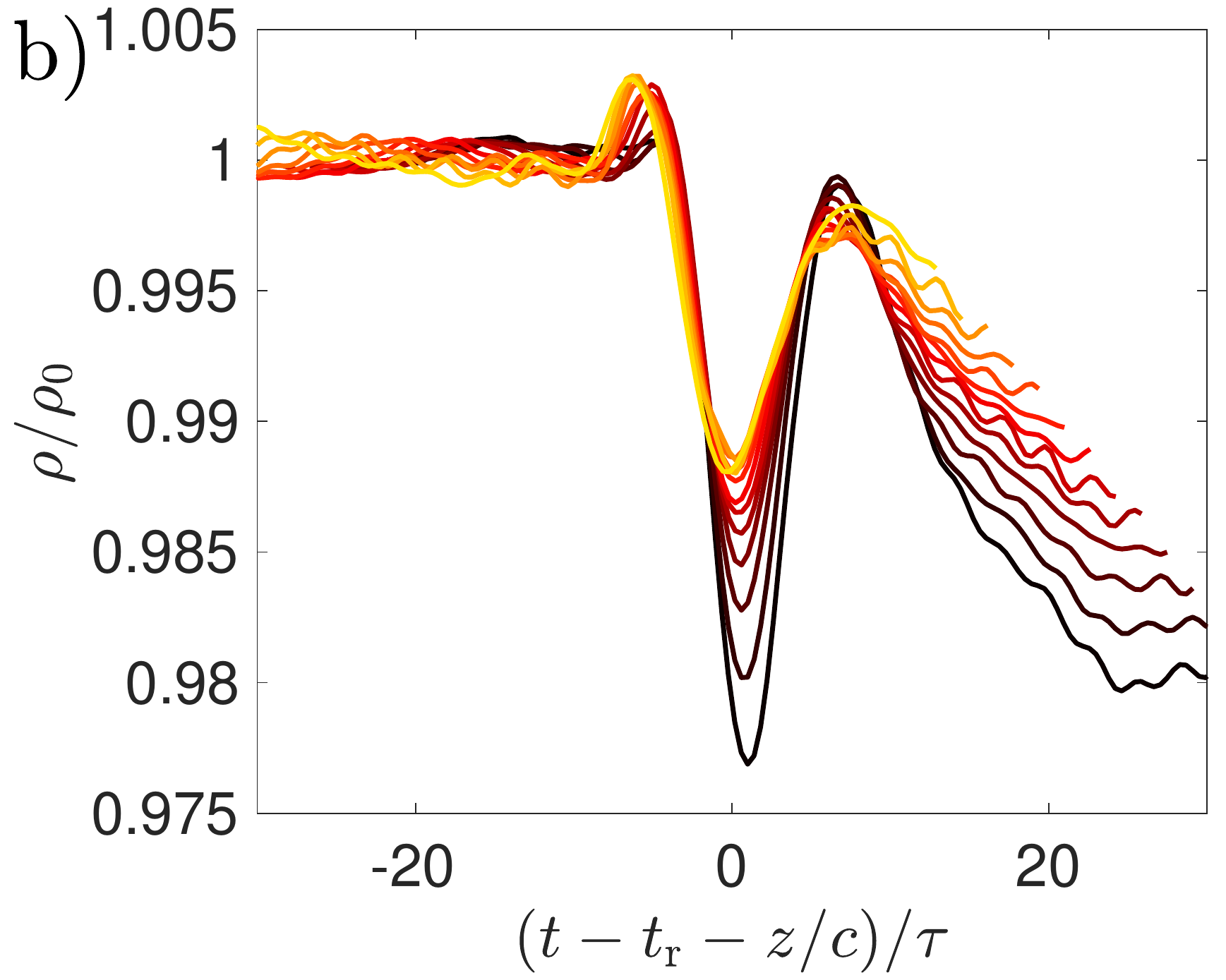} 
\includegraphics[width=0.3\textwidth]{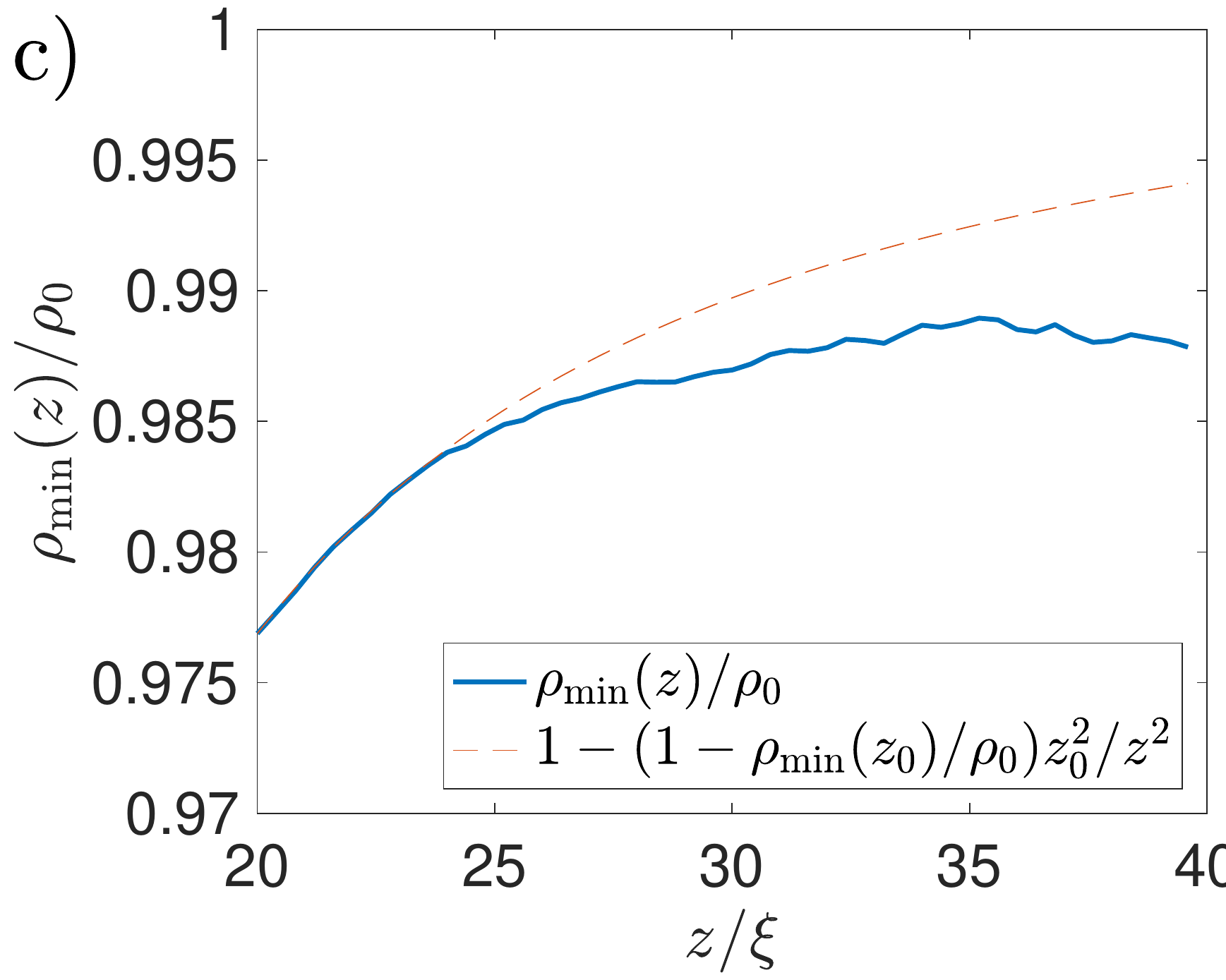} 
\caption{(Color online) Traveling pulse emitted during a reconnection with $A_{\rm r }=1.67$. \textbf{(a)} Density around the pulse as function of time for different values of the distance $z$ to the reconnection plane. \textbf{(b)} Same as \textbf{(b)}, but as a function of the retarded time. \textbf{(c)} Depth of the traveling pulse trough $\rho_{\rm min}$ at different values of $z$ compared with the $1/z^2$ decay law predicted form an acoustic pulse emitted from a point source. The theoretical formula is obtained by imposing the $1/z^2$ decay and matching the measured value of $\rho_{\rm min}$ at $z_0=20\xi$, the smallest value of $z$ at which the pulse was measured. The speed of sound is $c$ and $\tau=\xi/c$.}
\label{Fig:Pulse}
\end{figure}
We can observe that the pulse appears to move slightly slower than the speed of sound $ c $.
The signal shows some dispersive effects while time advances, an evidence that the pulse probably contains more than one Bogoliubov perturbation, perhaps including high wave numbers. Also, the depth of its trough $\rho_{\rm min}$ decays as the pulse propagates following the scaling $ \propto z^{-2} $ typical of a three-dimensional wave signal originating from a point source, as depicted in Fig.~\ref{Fig:Pulse}(c). 
It is however still unclear if the the pulse consists of a simple linear superposition of Bogoliubov perturbations, or if a fully nonlinear subsonic coherent structure, like a Robert--Jones solitary wave \cite{Jones:1982aa}, is also superimposed.  
A complete analysis on the spectrum of the sound pulse and the possible presence of coherent structures is left for future works.

Concerning the energy transferred from the vortices to the pulse, we present in Fig.\ref{fig:DeltaE}a-b the comparison between our theoretical prediction, Eq.~(\ref{eq:Ewav}), and the GP reconnection data obtained in \cite{Villois2020Irreversible} for different choices of $ \delta_{\rm lin} $.
\begin{figure}[h]
\centering{
\includegraphics[width=0.32\textwidth]{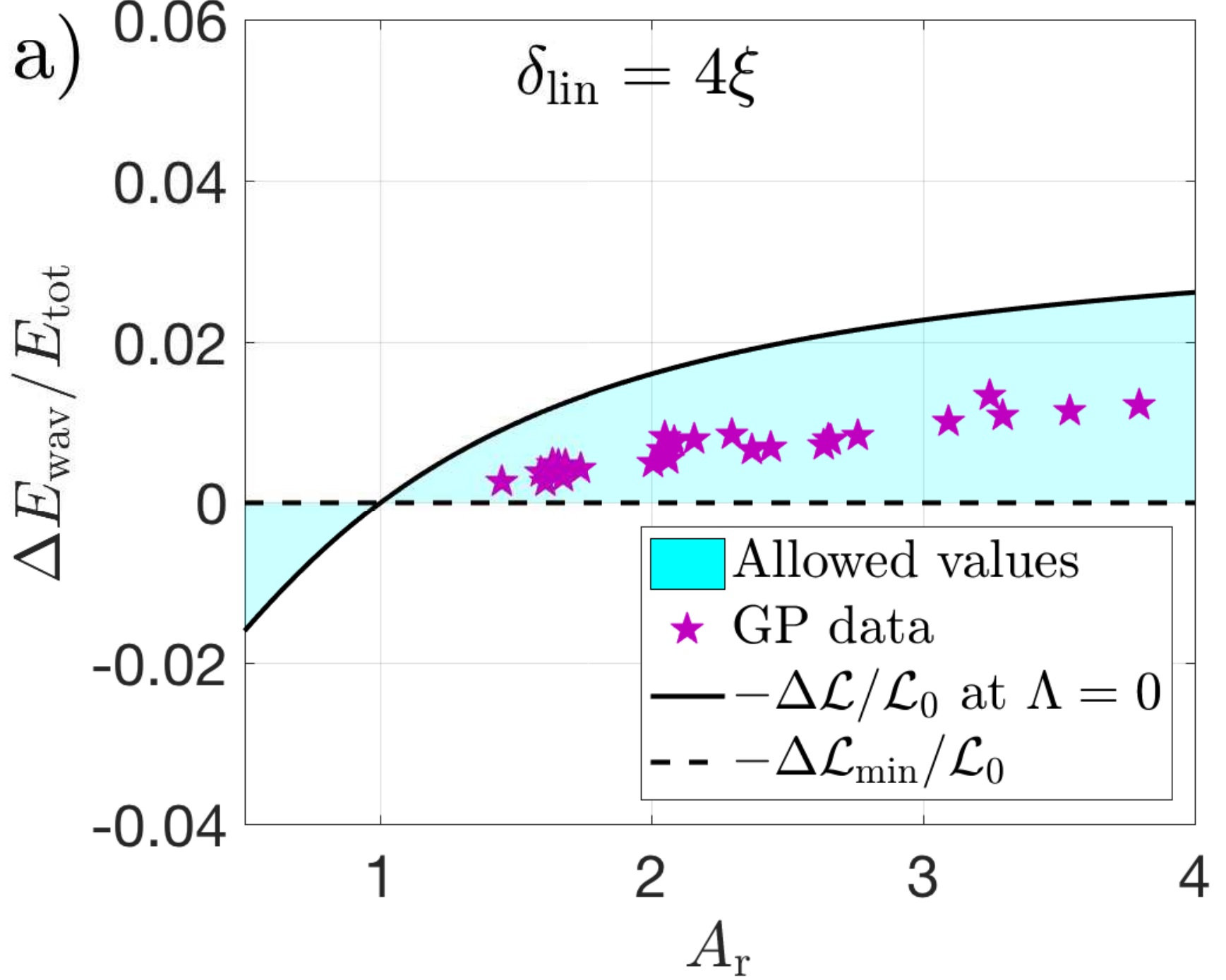} 
\includegraphics[width=0.32\textwidth]{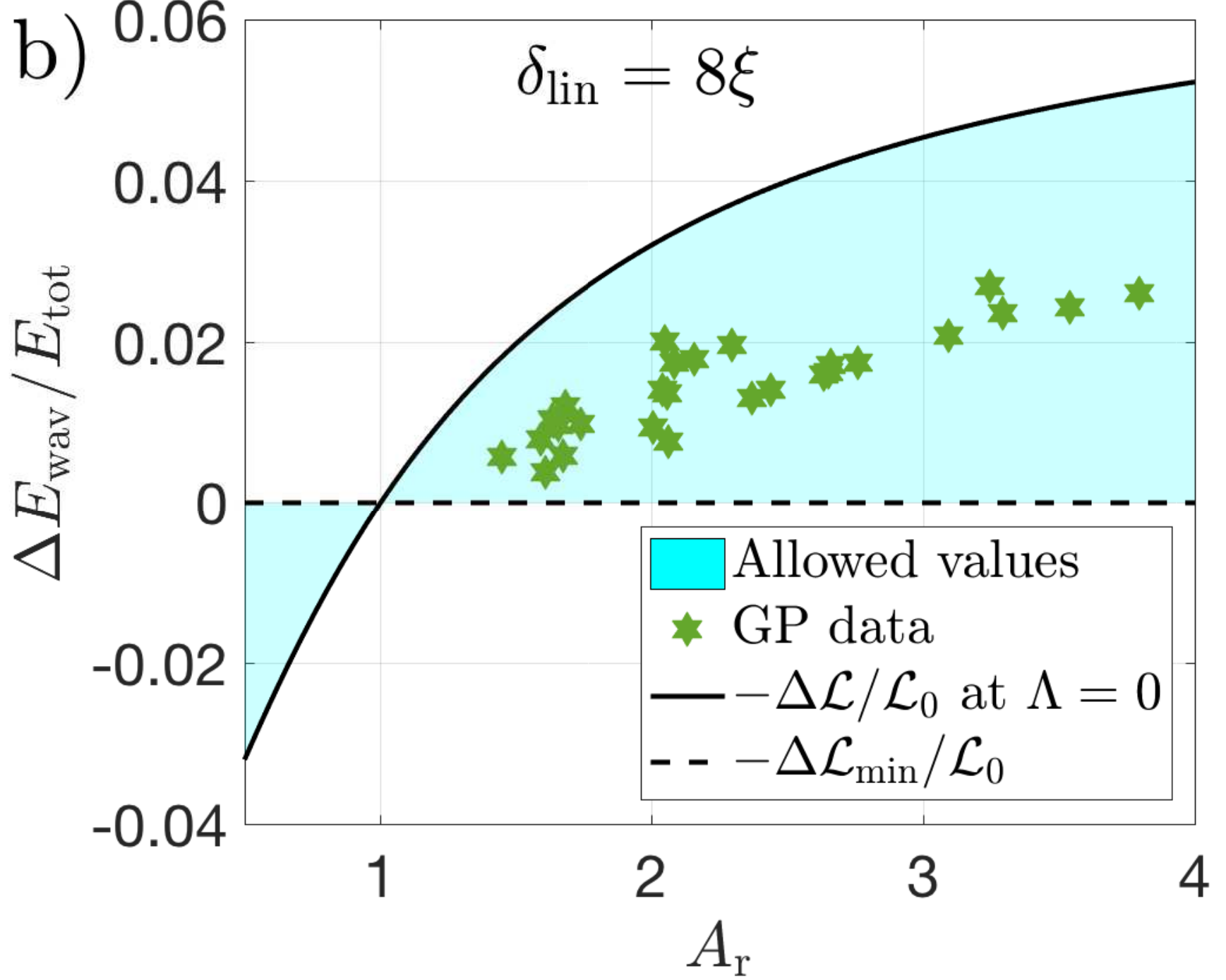} 
\includegraphics[width=0.32\textwidth]{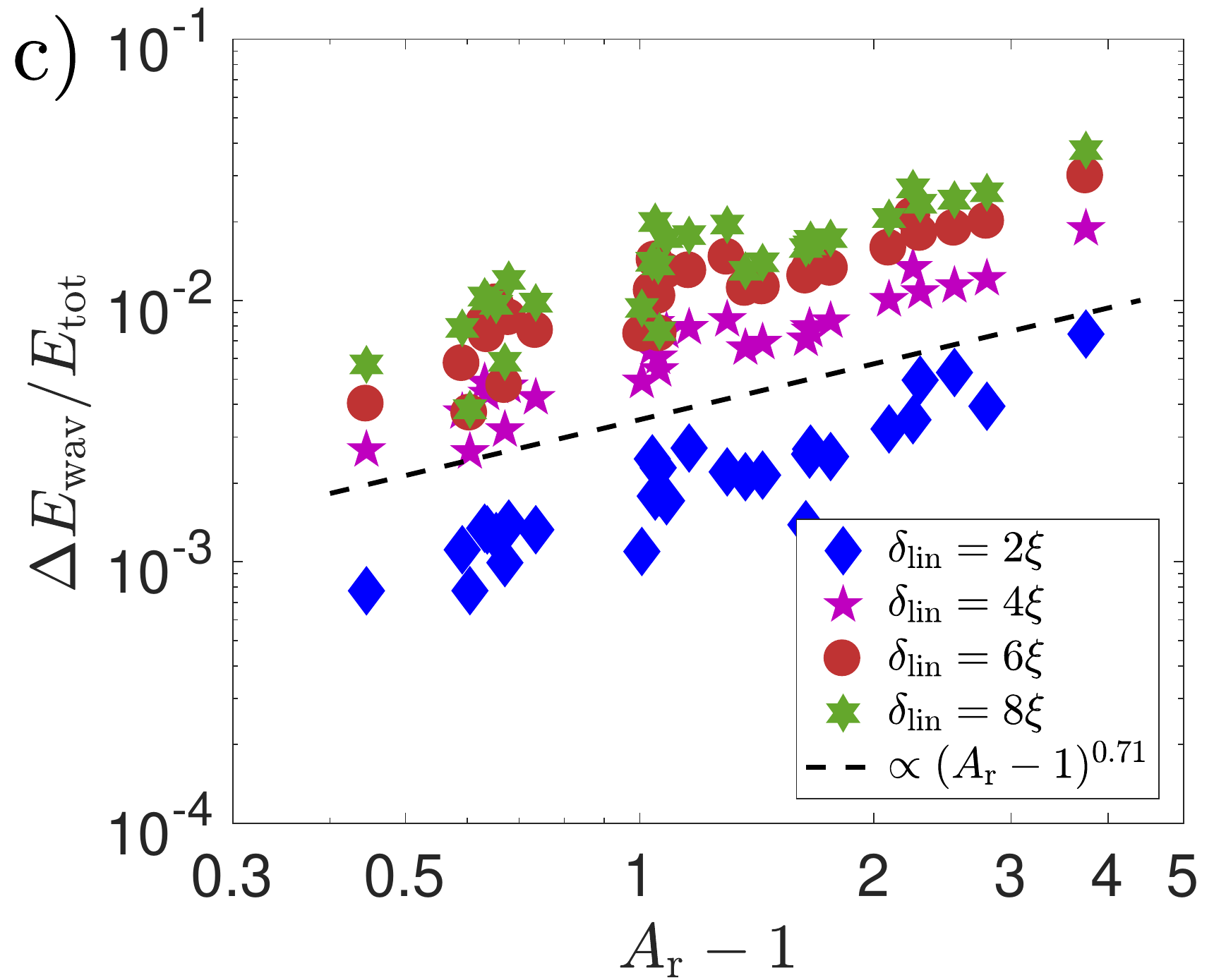} 
}
\caption{(Colour online) Energy radiated during reconnections as a function of  $A_{\rm r}=A^+/A^-$ for different values of $ \delta_{\rm lin}$. Numerical data from GP simulations obtained in \cite{Villois2020Irreversible} is confronted to our theoretical prediction in Eq.~(\ref{eq:Ewav}). \textbf{(a)} $ \delta_{\rm lin}=4 \xi $. \textbf{(b)} $ \delta_{\rm lin}=8 \xi $. \textbf{(c)} LogLog plot of energy as a function of $A_{\rm r}-1$ for different values of $ \delta_{\rm lin}$.}
\label{fig:DeltaE}
\end{figure}
The region of validity of the theory, the colored regions depicted in Fig.~\ref{fig:DeltaE}, represents all the accessible values of the concavity parameter $ \Lambda \in \mathbb{R} $, with boundaries obtained from Eqs~(\ref{eq:enDiffLambda0}) and (\ref{eq:enDiffLambdaInf}).
We note that the specific value of the energy difference in Eq.~(\ref{eq:Ewav}) depends on the choice of $ \delta_{\rm lin} $, but the conclusion that reconnection with $A_r<1$ are unlucky to occur, remains valid. Moreover, the fact that no energy exchange takes place for a symmetric reconnection $A_r=1$, suggests to plot $\Delta E_{\rm wav}$ versus $A_r-1$. Such plot is displayed in Fig.\ref{fig:DeltaE}.c, where the scaling $\Delta E_{\rm wav}\sim(A_r-1)^{0.71}$, obtained by fit, is clearly observed for at least one decade and all values of $\delta_{\rm lin}$. A more accurate theory, providing such scaling exponent is out of the scope of this work.

It is interesting to note that in a recent experimental work~\cite{svancaralamantia2019}, P.~Svancara and M.~La Mantia observed that in superfluid helium, the skewness of Langrangian velocity increments exhibits the opposite trend with respect to the one of classical flows, namely, in superfluids particles accelerate on average faster than they decelerate. The authors interpreted their findings as a consequence of quantum vortex reconnections because particles could probe individual vortex reconnections. This interpretation is consistent with the recent numerical calculations reported in \cite{GiuriatoReconnections}, where quantum vortex reconnections mediated by particles were studied. In that work, the authors observed that in addition to the momentum exchange between the vortices and sound waves, there is also a net transfer towards particles that incurs on a fast acceleration of particles at the reconnection event.

We should also remark on the possible role played by the center-line helicity $ \mathcal{H}_c $ during a reconnection event.
As first observed in \cite{Scheeler28102014PNASDavide}, the evolution of $ \mathcal{H}_c $ shows a sudden drop (but still being continuous) during a reconnection if the initial linking number between the vortex filaments is non-zero.
This is certainly the case analyzed in \cite{Villois2020Irreversible}, where the initial configuration of the filaments, an Hopf link, has linking number equal to 2. The temporal evolution of a Hopf link, for a case with $A_r=3.5$, is displayed Fig.~\ref{Fig:hel}.a, where the helicity drop is clearly visible.
One might be tempted to think that the sudden drop in the center-line helicity $ \Delta\mathcal{H}_c $ is related to the properties of the reconnecting filaments, for example their parameter $ A_{\rm r} $.
Figure~\ref{Fig:hel}.b shows that $ \Delta\mathcal{H}_c $ as a function of the parameter $A_{\rm r}$. 
\begin{figure}
\includegraphics[width=0.48\textwidth]{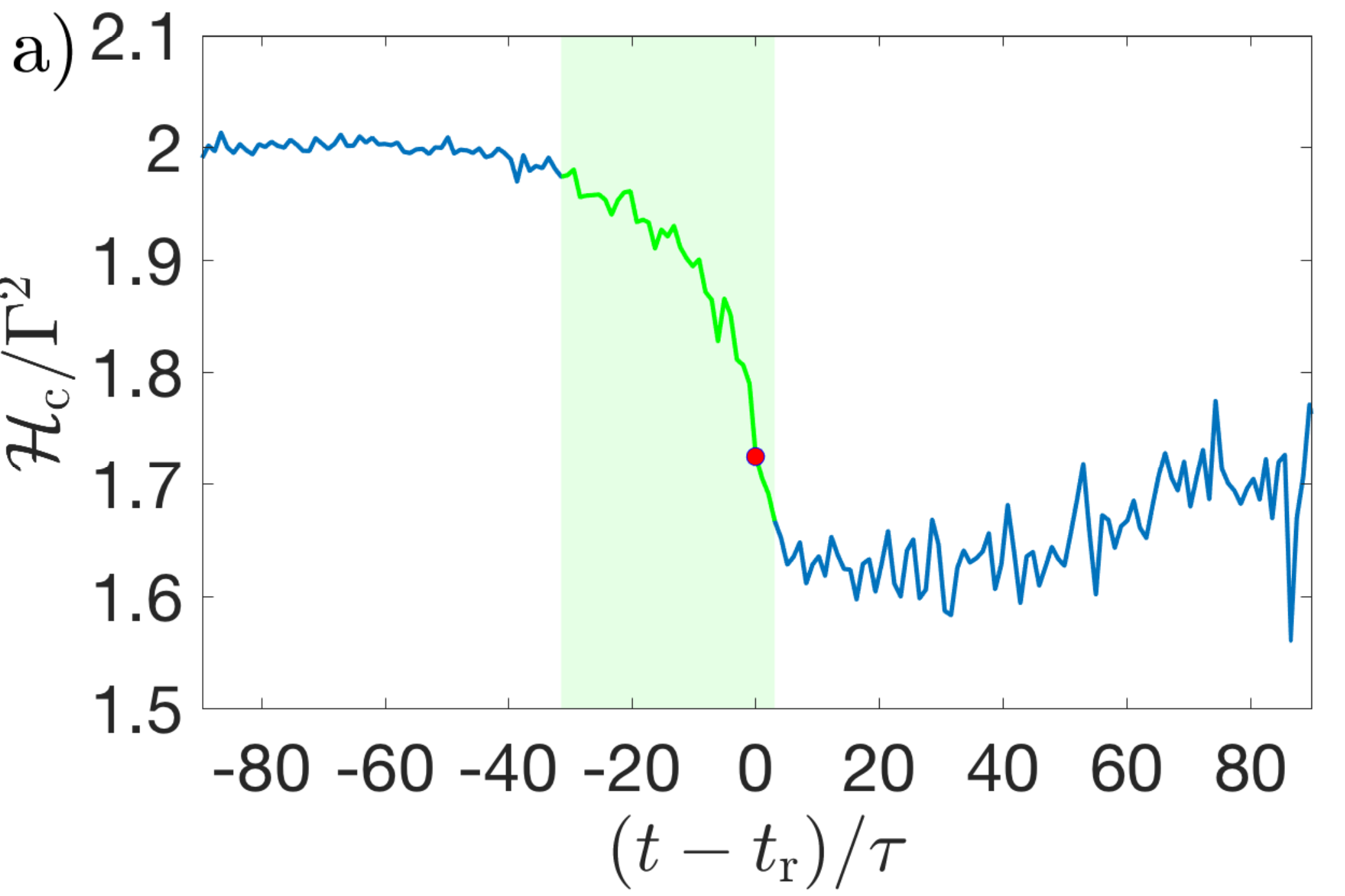} 
\includegraphics[width=0.48\textwidth]{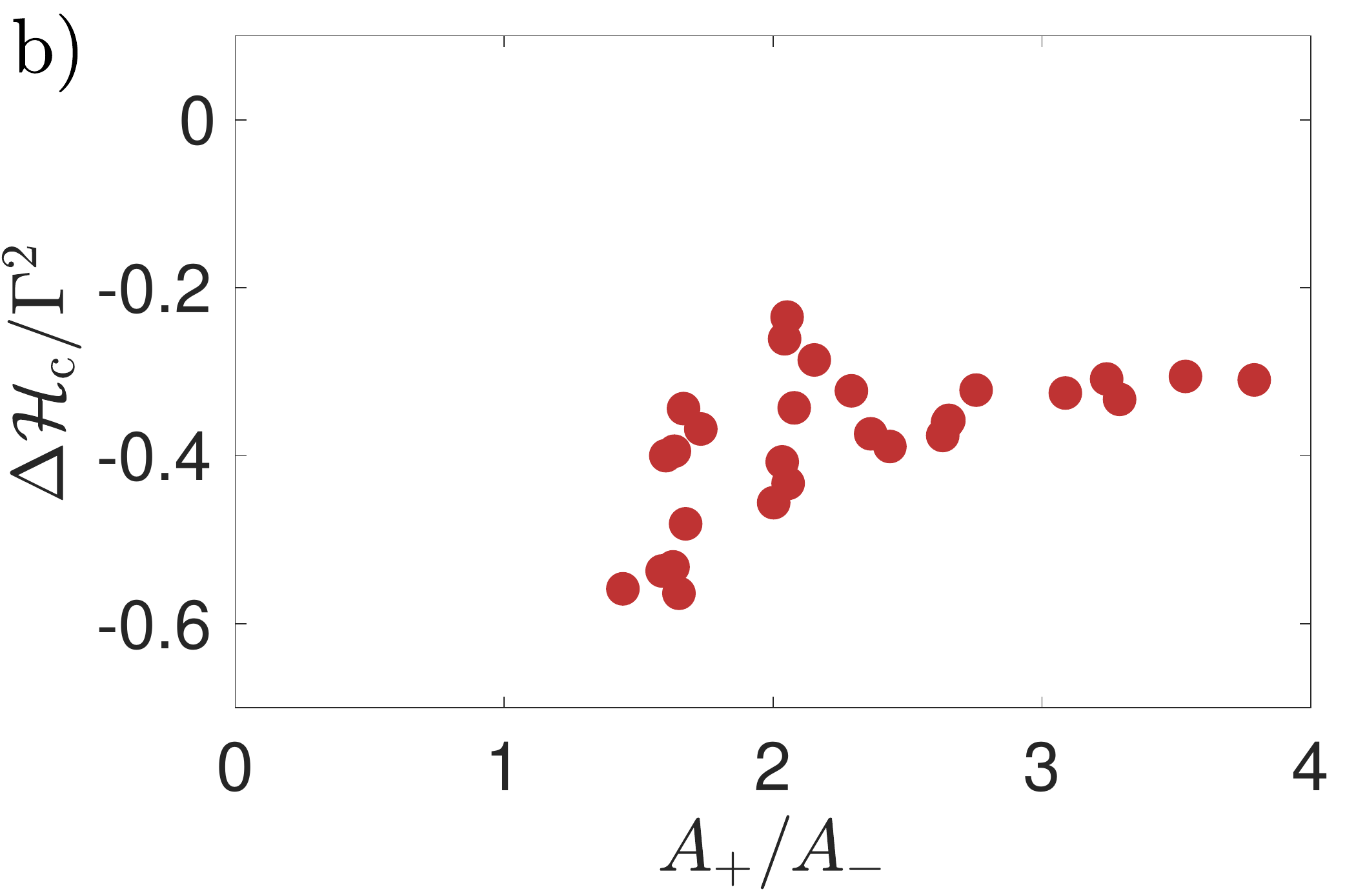} 
\caption{(Color online) \textbf{a)} Temporal evolution of central line helicity $H_{\rm c}$ for a reconnection having $A_{\rm r}=3.5$. The green zone correspond to the time interval defined by $\delta^\pm(t^\pm)=\delta_{\rm lin}=8\xi$. The red dot is the reconnection time. \textbf{a)} Drop of central line helicity $\Delta H_c=H_c(t^+)-H_c(t^-)$ as a function of $A_{\rm r}=A^+/A^-$. $\Gamma$ is the circulation of the filaments.
}
\label{Fig:hel}
\end{figure}
Numerical data do not correlate as well as the energy and no clear trend is observed.
We can however speculate that the center-line helicity drop may influence the amplitude and polarization of the Kelvin waves forming on the vortex filaments after the reconnection process.
This research proposal constitutes another interesting direction for future works.

In conclusion, our theoretical results help to understand not only {\it how} vortex reconnections take place in quantum fluids, but also {\it why} they do occur.
Albeit the GP model is time reversible, the vortex reconnection process presents a time asymmetry so that the system can naturally transfer part of the (kinetic) incompressible energy into its compressible counterpart. In some sense, this observed temporal asymmetry can be interpreted as consequence of the system being in an out-of-equilibrium stage and reconnections being a fast route towards reaching thermal equilibrium. It will be very interesting to study vortex reconnections at finite temperatures, where a thermal bath of Bogoliubov modes is present, to see if whether this asymmetry is reduced or destroyed completely.
Furthermore, the situation can be different when open conditions, like external forcing and damping terms acting at different length scales, are introduced. 
In a fully developed turbulent state, fluctuations could provide enough energy to allow reconnections with $A_r<1$, but one might still expect skew distributions towards $A_r>1$, as dissipation of turbulent flows do not vanishes in the limit of infinite Reynolds numbers because of the dissipative anomaly of turbulence \cite{Frisch1995}. This is undoubtedly, another interesting direction of research for future studies.

Finally, it is important to remark that the theoretical predictions obtained in this work may complement future experiments in several physical systems.
In Bose--Einstein condensates made of dilute gases, experimentalists are now able to observe the real-time dynamics of few reconnecting vortex filaments \cite{PhysRevLett.115.170402} and are now working to create reproducible reconnections \cite{Xhani:2020aa}.
Even if measuring the incompressible and compressible energy components of the superfluid is still a huge experimental challenge, assessing the directionality and shape of the sound pulse should be nowadays achievable using, for example, density imaging or Bragg spectroscopy.
In superfluid liquid helium-4 experiments, the challenge is to track single vortex reconnections, especially in the limit of very low temperatures where the normal component is negligible.
However, measuring the production of compressible energy excitations (and/or rotons) during many reconnections occurring in a turbulent tangle might be achievable \cite{PhysRevLett.121.015302, PhysRevE.85.036306, Scheeler28102014PNASDavide}; in order to be able to theoretically estimate this production rate, a statistical extension to our detailed analysis, including possibly finite temperature effects, might be needed.
Ultimately, let us point out there are impressive similarities between vortex reconnections occurring in quantum fluids and classical fluids at high Reynolds numbers, especially when the classical vortex tubes possess an almost hollow core \cite{Kleckner:2016aa, Alekseenko:2016aa}.
It would be therefore very interesting to measure the pre-factors $ A^\pm $, reconnection angle $ \phi^+ $ and concavity parameter $ \Lambda $ in the classical fluid experiments and quantitatively compare them with our theoretical predictions.

\appendix

\section{Choice of system of coordinates and value of parameter $\theta$}

To simplify the analysis of the wave-function parameters and the following calculations we have set the parameter $\theta=0$. Actually, one can always perform a rotation in the $x-y$ plane such that Eq.\eqref{eq:genICProj-xy-theta} becomes Eq.\eqref{eq:genICProj-xy} after a redefinition of the coefficient $A,B,C,D$. Once this change of variable is performed, Eq.\eqref{eq:nodal_im} will become in the new variables
\begin{equation}
  z = \frac{C_{xx} x^2 +C_{xy}xy +C_{yy} y^2}{2 \zeta} + \frac{A+B}{4 p \pi \zeta} \, \Gamma t \, ,
  \label{eq:nodal_imAppendix}
  \end{equation}
where $C_{xx},C_{xy}$ and $C_{yy}$ are new coefficient depending on the old ones. The analisys concerning the region of parameteres involves only the coefficients of the hyperbola, therefore it remains unchanged. The concavity parameter $\Lambda$ can not be defined so straighforwardly as now the projection on the $x=0$ or $y=0$ are not simple parabolas. However, the concavity increases with the magnitud of the coefficiens $C_{xx},C_{xy}$ and $C_{yy}$.

The momentum difference \eqref{eq:DeltaPz} remains unchanged if \eqref{eq:nodal_imAppendix} is taken into consideration for the vortex parametrization. Finally, the expression for the difference of vortex length are slightly more complicated, but the property that $\Delta\mathcal{L}\ge 0$ for $A_{\rm r}\le1$ and $\Delta\mathcal{L}\le 0$ for $A_{\rm r}\ge1$ remains valid.

\begin{acknowledgments} 
The authors acknowledge fruitful discussions with J. Hannay that led to the considerations of the most general normal form discussed in the Appendix of this manuscript.
The authors acknowledge L.~Galantucci for providing data displayed in Fig.\ref{Fig:ApAm} and A.~Villois for supplying the raw data used in Fig.\ref{Fig:Pulse}.
G.K., D.P. were supported by the cost-share Royal Society International Exchanges Scheme (IE150527) in conjunction with CNRS.
D.P. was supported by the EPSRC First Grant scheme (EP/P023770/1).  
G.K. and D.P. acknowledge the Federation Doeblin for supporting DP during his sojourn in Nice. 
G.K. was also supported by the ANR JCJC GIANTE ANR-18-CE30-0020-0.1 and by the EU Horizon 2020 research and innovation programme under the grant agreement No 823937 in the framework of Marie Skodowska-Curie HALT project.
Computations were carried out at M\'esocentre SIGAMM hosted at the Observatoire
de la C\^ote d'Azur and on the High Performance Computing Cluster supported by the Research and Specialist Computing Support service at the University of East Anglia.
Part of this work has been presented at the workshop ``Irreversibility and Turbulence''  hosted by Fondation Les Treilles in September 2017. 
The authors acknowledge Fondation Les Treilles and all participants of the workshop for the frightful scientific discussions and support.
D.P. acknowledges J.~Hannay for his frequent visits to Norfolk that always lead to fruitful conversations. 
\end{acknowledgments}


\begin{thebibliography}{0}%
\makeatletter
\providecommand \@ifxundefined [1]{%
 \@ifx{#1\undefined}
}%
\providecommand \@ifnum [1]{%
 \ifnum #1\expandafter \@firstoftwo
 \else \expandafter \@secondoftwo
 \fi
}%
\providecommand \@ifx [1]{%
 \ifx #1\expandafter \@firstoftwo
 \else \expandafter \@secondoftwo
 \fi
}%
\providecommand \natexlab [1]{#1}%
\providecommand \enquote  [1]{``#1''}%
\providecommand \bibnamefont  [1]{#1}%
\providecommand \bibfnamefont [1]{#1}%
\providecommand \citenamefont [1]{#1}%
\providecommand \href@noop [0]{\@secondoftwo}%
\providecommand \href [0]{\begingroup \@sanitize@url \@href}%
\providecommand \@href[1]{\@@startlink{#1}\@@href}%
\providecommand \@@href[1]{\endgroup#1\@@endlink}%
\providecommand \@sanitize@url [0]{\catcode `\\12\catcode `\$12\catcode
  `\&12\catcode `\#12\catcode `\^12\catcode `\_12\catcode `\%12\relax}%
\providecommand \@@startlink[1]{}%
\providecommand \@@endlink[0]{}%
\providecommand \url  [0]{\begingroup\@sanitize@url \@url }%
\providecommand \@url [1]{\endgroup\@href {#1}{\urlprefix }}%
\providecommand \urlprefix  [0]{URL }%
\providecommand \Eprint [0]{\href }%
\providecommand \doibase [0]{https://doi.org/}%
\providecommand \selectlanguage [0]{\@gobble}%
\providecommand \bibinfo  [0]{\@secondoftwo}%
\providecommand \bibfield  [0]{\@secondoftwo}%
\providecommand \translation [1]{[#1]}%
\providecommand \BibitemOpen [0]{}%
\providecommand \bibitemStop [0]{}%
\providecommand \bibitemNoStop [0]{.\EOS\space}%
\providecommand \EOS [0]{\spacefactor3000\relax}%
\providecommand \BibitemShut  [1]{\csname bibitem#1\endcsname}%
\let\auto@bib@innerbib\@empty
\end{thebibliography}%


\begin{thebibliography}{49}%
\makeatletter
\providecommand \@ifxundefined [1]{%
 \@ifx{#1\undefined}
}%
\providecommand \@ifnum [1]{%
 \ifnum #1\expandafter \@firstoftwo
 \else \expandafter \@secondoftwo
 \fi
}%
\providecommand \@ifx [1]{%
 \ifx #1\expandafter \@firstoftwo
 \else \expandafter \@secondoftwo
 \fi
}%
\providecommand \natexlab [1]{#1}%
\providecommand \enquote  [1]{``#1''}%
\providecommand \bibnamefont  [1]{#1}%
\providecommand \bibfnamefont [1]{#1}%
\providecommand \citenamefont [1]{#1}%
\providecommand \href@noop [0]{\@secondoftwo}%
\providecommand \href [0]{\begingroup \@sanitize@url \@href}%
\providecommand \@href[1]{\@@startlink{#1}\@@href}%
\providecommand \@@href[1]{\endgroup#1\@@endlink}%
\providecommand \@sanitize@url [0]{\catcode `\\12\catcode `\$12\catcode
  `\&12\catcode `\#12\catcode `\^12\catcode `\_12\catcode `\%12\relax}%
\providecommand \@@startlink[1]{}%
\providecommand \@@endlink[0]{}%
\providecommand \url  [0]{\begingroup\@sanitize@url \@url }%
\providecommand \@url [1]{\endgroup\@href {#1}{\urlprefix }}%
\providecommand \urlprefix  [0]{URL }%
\providecommand \Eprint [0]{\href }%
\providecommand \doibase [0]{https://doi.org/}%
\providecommand \selectlanguage [0]{\@gobble}%
\providecommand \bibinfo  [0]{\@secondoftwo}%
\providecommand \bibfield  [0]{\@secondoftwo}%
\providecommand \translation [1]{[#1]}%
\providecommand \BibitemOpen [0]{}%
\providecommand \bibitemStop [0]{}%
\providecommand \bibitemNoStop [0]{.\EOS\space}%
\providecommand \EOS [0]{\spacefactor3000\relax}%
\providecommand \BibitemShut  [1]{\csname bibitem#1\endcsname}%
\let\auto@bib@innerbib\@empty
\bibitem [{\citenamefont {Villois}\ \emph {et~al.}(2020)\citenamefont
  {Villois}, \citenamefont {Proment},\ and\ \citenamefont
  {Krstulovic}}]{Villois2020Irreversible}%
  \BibitemOpen
  \bibfield  {author} {\bibinfo {author} {\bibfnamefont {A.}~\bibnamefont
  {Villois}}, \bibinfo {author} {\bibfnamefont {D.}~\bibnamefont {Proment}},\
  and\ \bibinfo {author} {\bibfnamefont {G.}~\bibnamefont {Krstulovic}},\
  }\bibfield  {title} {\bibinfo {title} {Irreversible dynamics of vortex
  reconnections in quantum fluids},\ }\href@noop {} {\bibfield  {journal}
  {\bibinfo  {journal} {arXiv:2005.02048}\ } (\bibinfo {year}
  {2020})}\BibitemShut {NoStop}%
\bibitem [{\citenamefont {Saffman}(1993)}]{Saffman:1993aa}%
  \BibitemOpen
  \bibfield  {author} {\bibinfo {author} {\bibfnamefont {P.~G.}\ \bibnamefont
  {Saffman}},\ }\href {https://doi.org/DOI: 10.1017/CBO9780511624063} {\emph
  {\bibinfo {title} {Cambridge Monographs on Mechanics}}}\ (\bibinfo
  {publisher} {Cambridge University Press},\ \bibinfo {address} {Cambridge},\
  \bibinfo {year} {1993})\BibitemShut {NoStop}%
\bibitem [{\citenamefont {Priest}(1999)}]{Priest1999}%
  \BibitemOpen
  \bibfield  {author} {\bibinfo {author} {\bibfnamefont {E.~R.}\ \bibnamefont
  {Priest}},\ }\bibinfo {title} {Heating the solar corona by magnetic
  reconnection},\ in\ \href {https://doi.org/10.1007/978-94-011-4203-8_8}
  {\emph {\bibinfo {booktitle} {Plasma Astrophysics And Space Physics:
  Proceedings of the VIIth International Conference held in Lindau, Germany,
  May 4--8, 1998}}},\ \bibinfo {editor} {edited by\ \bibinfo {editor}
  {\bibfnamefont {J.}~\bibnamefont {B{\"u}chner}}, \bibinfo {editor}
  {\bibfnamefont {I.}~\bibnamefont {Axford}}, \bibinfo {editor} {\bibfnamefont
  {E.}~\bibnamefont {Marsch}},\ and\ \bibinfo {editor} {\bibfnamefont
  {V.}~\bibnamefont {Vasyli{\={u}}nas}}}\ (\bibinfo  {publisher} {Springer
  Netherlands},\ \bibinfo {address} {Dordrecht},\ \bibinfo {year} {1999})\ pp.\
  \bibinfo {pages} {77--100}\BibitemShut {NoStop}%
\bibitem [{\citenamefont {Kida}\ and\ \citenamefont
  {Takaoka}(1994)}]{Kida1994}%
  \BibitemOpen
  \bibfield  {author} {\bibinfo {author} {\bibfnamefont {S.}~\bibnamefont
  {Kida}}\ and\ \bibinfo {author} {\bibfnamefont {M.}~\bibnamefont {Takaoka}},\
  }\bibfield  {title} {\bibinfo {title} {Vortex reconnection},\ }\href
  {https://doi.org/10.1146/annurev.fl.26.010194.001125} {\bibfield  {journal}
  {\bibinfo  {journal} {Annual Review of Fluid Mechanics}\ }\textbf {\bibinfo
  {volume} {26}},\ \bibinfo {pages} {169} (\bibinfo {year} {1994})},\ \Eprint
  {https://arxiv.org/abs/http://dx.doi.org/10.1146/annurev.fl.26.010194.001125}
  {http://dx.doi.org/10.1146/annurev.fl.26.010194.001125} \BibitemShut
  {NoStop}%
\bibitem [{\citenamefont {Fonda}\ \emph {et~al.}(2014)\citenamefont {Fonda},
  \citenamefont {Meichle}, \citenamefont {Ouellette}, \citenamefont {Hormoz},\
  and\ \citenamefont {Lathrop}}]{Fonda25032014}%
  \BibitemOpen
  \bibfield  {author} {\bibinfo {author} {\bibfnamefont {E.}~\bibnamefont
  {Fonda}}, \bibinfo {author} {\bibfnamefont {D.~P.}\ \bibnamefont {Meichle}},
  \bibinfo {author} {\bibfnamefont {N.~T.}\ \bibnamefont {Ouellette}}, \bibinfo
  {author} {\bibfnamefont {S.}~\bibnamefont {Hormoz}},\ and\ \bibinfo {author}
  {\bibfnamefont {D.~P.}\ \bibnamefont {Lathrop}},\ }\bibfield  {title}
  {\bibinfo {title} {Direct observation of kelvin waves excited by quantized
  vortex reconnection},\ }\href {https://doi.org/10.1073/pnas.1312536110}
  {\bibfield  {journal} {\bibinfo  {journal} {Proceedings of the National
  Academy of Sciences}\ }\textbf {\bibinfo {volume} {111}},\ \bibinfo {pages}
  {4707} (\bibinfo {year} {2014})}\BibitemShut {NoStop}%
\bibitem [{\citenamefont {Serafini}\ \emph {et~al.}(2017)\citenamefont
  {Serafini}, \citenamefont {Galantucci}, \citenamefont {Iseni}, \citenamefont
  {Bienaim{\'e}}, \citenamefont {Bisset}, \citenamefont {Barenghi},
  \citenamefont {Dalfovo}, \citenamefont {Lamporesi},\ and\ \citenamefont
  {Ferrari}}]{Serafini:2017aa}%
  \BibitemOpen
  \bibfield  {author} {\bibinfo {author} {\bibfnamefont {S.}~\bibnamefont
  {Serafini}}, \bibinfo {author} {\bibfnamefont {L.}~\bibnamefont
  {Galantucci}}, \bibinfo {author} {\bibfnamefont {E.}~\bibnamefont {Iseni}},
  \bibinfo {author} {\bibfnamefont {T.}~\bibnamefont {Bienaim{\'e}}}, \bibinfo
  {author} {\bibfnamefont {R.~N.}\ \bibnamefont {Bisset}}, \bibinfo {author}
  {\bibfnamefont {C.~F.}\ \bibnamefont {Barenghi}}, \bibinfo {author}
  {\bibfnamefont {F.}~\bibnamefont {Dalfovo}}, \bibinfo {author} {\bibfnamefont
  {G.}~\bibnamefont {Lamporesi}},\ and\ \bibinfo {author} {\bibfnamefont
  {G.}~\bibnamefont {Ferrari}},\ }\bibfield  {title} {\bibinfo {title} {Vortex
  reconnections and rebounds in trapped atomic bose-einstein condensates},\
  }\href {https://doi.org/10.1103/PhysRevX.7.021031} {\bibfield  {journal}
  {\bibinfo  {journal} {Physical Review X}\ }\textbf {\bibinfo {volume} {7}},\
  \bibinfo {pages} {021031} (\bibinfo {year} {2017})}\BibitemShut {NoStop}%
\bibitem [{\citenamefont {Hussain}\ and\ \citenamefont
  {Duraisamy}(2011)}]{Fazle&KarthikPof2011}%
  \BibitemOpen
  \bibfield  {author} {\bibinfo {author} {\bibfnamefont {F.}~\bibnamefont
  {Hussain}}\ and\ \bibinfo {author} {\bibfnamefont {K.}~\bibnamefont
  {Duraisamy}},\ }\bibfield  {title} {\bibinfo {title} {Mechanics of viscous
  vortex reconnection},\ }\href
  {https://doi.org/http://dx.doi.org/10.1063/1.3532039} {\bibfield  {journal}
  {\bibinfo  {journal} {Physics of Fluids}\ }\textbf {\bibinfo {volume} {23}},\
  \bibinfo {eid} {021701} (\bibinfo {year} {2011})}\BibitemShut {NoStop}%
\bibitem [{\citenamefont {{Xue Zhike}}\ \emph {et~al.}(2016)\citenamefont {{Xue
  Zhike}}, \citenamefont {{Yan Xiaoli}}, \citenamefont {{Cheng Xin}},
  \citenamefont {{Yang Liheng}}, \citenamefont {{Su Yingna}}, \citenamefont
  {{Kliem Bernhard}}, \citenamefont {{Zhang Jun}}, \citenamefont {{Liu Zhong}},
  \citenamefont {{Bi Yi}}, \citenamefont {{Xiang Yongyuan}}, \citenamefont
  {{Yang Kai}},\ and\ \citenamefont {{Zhao Li}}}]{Zhike2016}%
  \BibitemOpen
  \bibfield  {author} {\bibinfo {author} {\bibnamefont {{Xue Zhike}}}, \bibinfo
  {author} {\bibnamefont {{Yan Xiaoli}}}, \bibinfo {author} {\bibnamefont
  {{Cheng Xin}}}, \bibinfo {author} {\bibnamefont {{Yang Liheng}}}, \bibinfo
  {author} {\bibnamefont {{Su Yingna}}}, \bibinfo {author} {\bibnamefont
  {{Kliem Bernhard}}}, \bibinfo {author} {\bibnamefont {{Zhang Jun}}}, \bibinfo
  {author} {\bibnamefont {{Liu Zhong}}}, \bibinfo {author} {\bibnamefont {{Bi
  Yi}}}, \bibinfo {author} {\bibnamefont {{Xiang Yongyuan}}}, \bibinfo {author}
  {\bibnamefont {{Yang Kai}}},\ and\ \bibinfo {author} {\bibnamefont {{Zhao
  Li}}},\ }\bibfield  {title} {\bibinfo {title} {{Observing the release of
  twist by magnetic reconnection in a solar filament eruption}},\ }\bibfield
  {journal} {\bibinfo  {journal} {Nat Commun}\ }\textbf {\bibinfo {volume}
  {7}},\ \href {https://doi.org/http://dx.doi.org/10.1038/ncomms11837
  10.1038/ncomms11837} {http://dx.doi.org/10.1038/ncomms11837
  10.1038/ncomms11837} (\bibinfo {year} {2016})\BibitemShut {NoStop}%
\bibitem [{\citenamefont {Yao}\ and\ \citenamefont
  {Hussain}(2020)}]{Yao:2020aa}%
  \BibitemOpen
  \bibfield  {author} {\bibinfo {author} {\bibfnamefont {J.}~\bibnamefont
  {Yao}}\ and\ \bibinfo {author} {\bibfnamefont {F.}~\bibnamefont {Hussain}},\
  }\bibfield  {title} {\bibinfo {title} {A physical model of turbulence cascade
  via vortex reconnection sequence and avalanche},\ }\href
  {https://doi.org/DOI: 10.1017/jfm.2019.905} {\bibfield  {journal} {\bibinfo
  {journal} {Journal of Fluid Mechanics}\ }\textbf {\bibinfo {volume} {883}},\
  \bibinfo {pages} {A51} (\bibinfo {year} {2020})}\BibitemShut {NoStop}%
\bibitem [{\citenamefont {Pumir}\ and\ \citenamefont
  {Kerr}(1987)}]{Pumir:1987aa}%
  \BibitemOpen
  \bibfield  {author} {\bibinfo {author} {\bibfnamefont {A.}~\bibnamefont
  {Pumir}}\ and\ \bibinfo {author} {\bibfnamefont {R.~M.}\ \bibnamefont
  {Kerr}},\ }\bibfield  {title} {\bibinfo {title} {Numerical simulation of
  interacting vortex tubes},\ }\href
  {https://doi.org/10.1103/PhysRevLett.58.1636} {\bibfield  {journal} {\bibinfo
   {journal} {Physical Review Letters}\ }\textbf {\bibinfo {volume} {58}},\
  \bibinfo {pages} {1636} (\bibinfo {year} {1987})}\BibitemShut {NoStop}%
\bibitem [{\citenamefont {Agafontsev}\ \emph {et~al.}(2018)\citenamefont
  {Agafontsev}, \citenamefont {Kuznetsov},\ and\ \citenamefont
  {Mailybaev}}]{Agafontsev:2018aa}%
  \BibitemOpen
  \bibfield  {author} {\bibinfo {author} {\bibfnamefont {D.~S.}\ \bibnamefont
  {Agafontsev}}, \bibinfo {author} {\bibfnamefont {E.~A.}\ \bibnamefont
  {Kuznetsov}},\ and\ \bibinfo {author} {\bibfnamefont {A.~A.}\ \bibnamefont
  {Mailybaev}},\ }\bibfield  {title} {\bibinfo {title} {Development of high
  vorticity structures and geometrical properties of the vortex line
  representation},\ }\bibfield  {booktitle} {\emph {\bibinfo {booktitle}
  {Physics of Fluids}},\ }\href {https://doi.org/10.1063/1.5049119} {\bibfield
  {journal} {\bibinfo  {journal} {Physics of Fluids}\ }\textbf {\bibinfo
  {volume} {30}},\ \bibinfo {pages} {095104} (\bibinfo {year}
  {2018})}\BibitemShut {NoStop}%
\bibitem [{\citenamefont {Kerr}(2013)}]{Kerr:2013aa}%
  \BibitemOpen
  \bibfield  {author} {\bibinfo {author} {\bibfnamefont {R.~M.}\ \bibnamefont
  {Kerr}},\ }\bibfield  {title} {\bibinfo {title} {Swirling, turbulent vortex
  rings formed from a chain reaction of reconnection events},\ }\bibfield
  {booktitle} {\emph {\bibinfo {booktitle} {Physics of Fluids}},\ }\href
  {https://doi.org/10.1063/1.4807060} {\bibfield  {journal} {\bibinfo
  {journal} {Physics of Fluids}\ }\textbf {\bibinfo {volume} {25}},\ \bibinfo
  {pages} {065101} (\bibinfo {year} {2013})}\BibitemShut {NoStop}%
\bibitem [{\citenamefont {McKeown}\ \emph {et~al.}(2020)\citenamefont
  {McKeown}, \citenamefont {Ostilla-M{\'o}nico}, \citenamefont {Pumir},
  \citenamefont {Brenner},\ and\ \citenamefont {Rubinstein}}]{McKeown:2020aa}%
  \BibitemOpen
  \bibfield  {author} {\bibinfo {author} {\bibfnamefont {R.}~\bibnamefont
  {McKeown}}, \bibinfo {author} {\bibfnamefont {R.}~\bibnamefont
  {Ostilla-M{\'o}nico}}, \bibinfo {author} {\bibfnamefont {A.}~\bibnamefont
  {Pumir}}, \bibinfo {author} {\bibfnamefont {M.~P.}\ \bibnamefont {Brenner}},\
  and\ \bibinfo {author} {\bibfnamefont {S.~M.}\ \bibnamefont {Rubinstein}},\
  }\bibfield  {title} {\bibinfo {title} {Turbulence generation through an
  iterative cascade of the elliptical instability},\ }\href
  {https://doi.org/10.1126/sciadv.aaz2717} {\bibfield  {journal} {\bibinfo
  {journal} {Science Advances}\ }\textbf {\bibinfo {volume} {6}},\ \bibinfo
  {pages} {eaaz2717} (\bibinfo {year} {2020})}\BibitemShut {NoStop}%
\bibitem [{\citenamefont {Pitaevskii}\ and\ \citenamefont
  {Stringari}(2016)}]{Pitaevskii:2016aa}%
  \BibitemOpen
  \bibfield  {author} {\bibinfo {author} {\bibfnamefont {L.}~\bibnamefont
  {Pitaevskii}}\ and\ \bibinfo {author} {\bibfnamefont {S.}~\bibnamefont
  {Stringari}},\ }\href@noop {} {\emph {\bibinfo {title} {Bose-Einstein
  condensation and superfluidity}}},\ Vol.\ \bibinfo {volume} {164 
  0191076686}\ (\bibinfo  {publisher} {Oxford University Press},\ \bibinfo
  {year} {2016})\BibitemShut {NoStop}%
\bibitem [{Note1()}]{Note1}%
  \BibitemOpen
  \bibinfo {note} {This is true for mean-field models of quantum fluids, when
  quantum fluctuations are considered this picture is more
  complicated.}\BibitemShut {Stop}%
\bibitem [{\citenamefont {Koplik}\ and\ \citenamefont
  {Levine}(1993)}]{KoplikLevinPRL1993}%
  \BibitemOpen
  \bibfield  {author} {\bibinfo {author} {\bibfnamefont {J.}~\bibnamefont
  {Koplik}}\ and\ \bibinfo {author} {\bibfnamefont {H.}~\bibnamefont
  {Levine}},\ }\bibfield  {title} {\bibinfo {title} {Vortex reconnection in
  superfluid helium},\ }\href {https://doi.org/10.1103/PhysRevLett.71.1375}
  {\bibfield  {journal} {\bibinfo  {journal} {Phys. Rev. Lett.}\ }\textbf
  {\bibinfo {volume} {71}},\ \bibinfo {pages} {1375} (\bibinfo {year}
  {1993})}\BibitemShut {NoStop}%
\bibitem [{\citenamefont {Nore}\ \emph {et~al.}(1997)\citenamefont {Nore},
  \citenamefont {Abid},\ and\ \citenamefont {Brachet}}]{nore1997decaying}%
  \BibitemOpen
  \bibfield  {author} {\bibinfo {author} {\bibfnamefont {C.}~\bibnamefont
  {Nore}}, \bibinfo {author} {\bibfnamefont {M.}~\bibnamefont {Abid}},\ and\
  \bibinfo {author} {\bibfnamefont {M.}~\bibnamefont {Brachet}},\ }\bibfield
  {title} {\bibinfo {title} {Decaying kolmogorov turbulence in a model of
  superflow},\ }\href@noop {} {\bibfield  {journal} {\bibinfo  {journal}
  {Physics of Fluids (1994-present)}\ }\textbf {\bibinfo {volume} {9}},\
  \bibinfo {pages} {2644} (\bibinfo {year} {1997})}\BibitemShut {NoStop}%
\bibitem [{\citenamefont {Villois}\ \emph
  {et~al.}(2016{\natexlab{a}})\citenamefont {Villois}, \citenamefont
  {Krstulovic}, \citenamefont {Proment},\ and\ \citenamefont
  {Salman}}]{VilloisTrackingAlgo}%
  \BibitemOpen
  \bibfield  {author} {\bibinfo {author} {\bibfnamefont {A.}~\bibnamefont
  {Villois}}, \bibinfo {author} {\bibfnamefont {G.}~\bibnamefont {Krstulovic}},
  \bibinfo {author} {\bibfnamefont {D.}~\bibnamefont {Proment}},\ and\ \bibinfo
  {author} {\bibfnamefont {H.}~\bibnamefont {Salman}},\ }\bibfield  {title}
  {\bibinfo {title} {A vortex filament tracking method for the
  gross--pitaevskii model of a superfluid},\ }\href
  {http://stacks.iop.org/1751-8121/49/i=41/a=415502} {\bibfield  {journal}
  {\bibinfo  {journal} {Journal of Physics A: Mathematical and Theoretical}\
  }\textbf {\bibinfo {volume} {49}},\ \bibinfo {pages} {415502} (\bibinfo
  {year} {2016}{\natexlab{a}})}\BibitemShut {NoStop}%
\bibitem [{\citenamefont {Villois}\ \emph
  {et~al.}(2016{\natexlab{b}})\citenamefont {Villois}, \citenamefont
  {Proment},\ and\ \citenamefont {Krstulovic}}]{PhysRevE.93.061103}%
  \BibitemOpen
  \bibfield  {author} {\bibinfo {author} {\bibfnamefont {A.}~\bibnamefont
  {Villois}}, \bibinfo {author} {\bibfnamefont {D.}~\bibnamefont {Proment}},\
  and\ \bibinfo {author} {\bibfnamefont {G.}~\bibnamefont {Krstulovic}},\
  }\bibfield  {title} {\bibinfo {title} {Evolution of a superfluid vortex
  filament tangle driven by the gross-pitaevskii equation},\ }\href
  {https://doi.org/10.1103/PhysRevE.93.061103} {\bibfield  {journal} {\bibinfo
  {journal} {Phys. Rev. E}\ }\textbf {\bibinfo {volume} {93}},\ \bibinfo
  {pages} {061103(R)} (\bibinfo {year} {2016}{\natexlab{b}})}\BibitemShut
  {NoStop}%
\bibitem [{\citenamefont {Bustamante}\ and\ \citenamefont
  {Nazarenko}(2015)}]{BustamanteNazarenko}%
  \BibitemOpen
  \bibfield  {author} {\bibinfo {author} {\bibfnamefont {M.~D.}\ \bibnamefont
  {Bustamante}}\ and\ \bibinfo {author} {\bibfnamefont {S.}~\bibnamefont
  {Nazarenko}},\ }\bibfield  {title} {\bibinfo {title} {Derivation of the
  biot-savart equation from the nonlinear schr\"odinger equation},\ }\href
  {https://doi.org/10.1103/PhysRevE.92.053019} {\bibfield  {journal} {\bibinfo
  {journal} {Phys. Rev. E}\ }\textbf {\bibinfo {volume} {92}},\ \bibinfo
  {pages} {053019} (\bibinfo {year} {2015})}\BibitemShut {NoStop}%
\bibitem [{\citenamefont {Feynman}(1955)}]{Feynman195517}%
  \BibitemOpen
  \bibfield  {author} {\bibinfo {author} {\bibfnamefont {R.}~\bibnamefont
  {Feynman}},\ }\bibfield  {title} {\bibinfo {title} {Chapter \{II\}
  application of quantum mechanics to liquid helium}\ }(\bibinfo  {publisher}
  {Elsevier},\ \bibinfo {year} {1955})\ pp.\ \bibinfo {pages} {17 --
  53}\BibitemShut {NoStop}%
\bibitem [{\citenamefont {Nazarenko}\ and\ \citenamefont
  {West}(2003)}]{nazarenko2003analytical}%
  \BibitemOpen
  \bibfield  {author} {\bibinfo {author} {\bibfnamefont {S.}~\bibnamefont
  {Nazarenko}}\ and\ \bibinfo {author} {\bibfnamefont {R.}~\bibnamefont
  {West}},\ }\bibfield  {title} {\bibinfo {title} {Analytical solution for
  nonlinear schr{\"o}dinger vortex reconnection},\ }\href@noop {} {\bibfield
  {journal} {\bibinfo  {journal} {Journal of low temperature physics}\ }\textbf
  {\bibinfo {volume} {132}},\ \bibinfo {pages} {1} (\bibinfo {year}
  {2003})}\BibitemShut {NoStop}%
\bibitem [{\citenamefont {Kursa}\ \emph {et~al.}(2011)\citenamefont {Kursa},
  \citenamefont {Bajer},\ and\ \citenamefont
  {Lipniacki}}]{Kursa&Bajer&LipniackiPRB2011}%
  \BibitemOpen
  \bibfield  {author} {\bibinfo {author} {\bibfnamefont {M.}~\bibnamefont
  {Kursa}}, \bibinfo {author} {\bibfnamefont {K.}~\bibnamefont {Bajer}},\ and\
  \bibinfo {author} {\bibfnamefont {T.}~\bibnamefont {Lipniacki}},\ }\bibfield
  {title} {\bibinfo {title} {Cascade of vortex loops initiated by a single
  reconnection of quantum vortices},\ }\href
  {https://doi.org/10.1103/PhysRevB.83.014515} {\bibfield  {journal} {\bibinfo
  {journal} {Phys. Rev. B}\ }\textbf {\bibinfo {volume} {83}},\ \bibinfo
  {pages} {014515} (\bibinfo {year} {2011})}\BibitemShut {NoStop}%
\bibitem [{\citenamefont {Zuccher}\ \emph {et~al.}(2012)\citenamefont
  {Zuccher}, \citenamefont {Caliari}, \citenamefont {Baggaley},\ and\
  \citenamefont {Barenghi}}]{Zuccher&Caliari&Baggaley&BarenghiPof2012}%
  \BibitemOpen
  \bibfield  {author} {\bibinfo {author} {\bibfnamefont {S.}~\bibnamefont
  {Zuccher}}, \bibinfo {author} {\bibfnamefont {M.}~\bibnamefont {Caliari}},
  \bibinfo {author} {\bibfnamefont {A.~W.}\ \bibnamefont {Baggaley}},\ and\
  \bibinfo {author} {\bibfnamefont {C.~F.}\ \bibnamefont {Barenghi}},\
  }\bibfield  {title} {\bibinfo {title} {Quantum vortex reconnections},\ }\href
  {https://doi.org/http://dx.doi.org/10.1063/1.4772198} {\bibfield  {journal}
  {\bibinfo  {journal} {Physics of Fluids}\ }\textbf {\bibinfo {volume} {24}},\
  \bibinfo {eid} {125108} (\bibinfo {year} {2012})}\BibitemShut {NoStop}%
\bibitem [{\citenamefont {Villois}\ \emph {et~al.}(2017)\citenamefont
  {Villois}, \citenamefont {Proment},\ and\ \citenamefont
  {Krstulovic}}]{villoisPRF2018}%
  \BibitemOpen
  \bibfield  {author} {\bibinfo {author} {\bibfnamefont {A.}~\bibnamefont
  {Villois}}, \bibinfo {author} {\bibfnamefont {D.}~\bibnamefont {Proment}},\
  and\ \bibinfo {author} {\bibfnamefont {G.}~\bibnamefont {Krstulovic}},\
  }\bibfield  {title} {\bibinfo {title} {Universal and nonuniversal aspects of
  vortex reconnections in superfluids},\ }\href
  {https://doi.org/10.1103/PhysRevFluids.2.044701} {\bibfield  {journal}
  {\bibinfo  {journal} {Phys. Rev. Fluids}\ }\textbf {\bibinfo {volume} {2}},\
  \bibinfo {pages} {044701} (\bibinfo {year} {2017})}\BibitemShut {NoStop}%
\bibitem [{\citenamefont {Rorai}\ \emph {et~al.}(2016)\citenamefont {Rorai},
  \citenamefont {Skipper}, \citenamefont {Kerr},\ and\ \citenamefont
  {Sreenivasan}}]{Rorai:2016aa}%
  \BibitemOpen
  \bibfield  {author} {\bibinfo {author} {\bibfnamefont {C.}~\bibnamefont
  {Rorai}}, \bibinfo {author} {\bibfnamefont {J.}~\bibnamefont {Skipper}},
  \bibinfo {author} {\bibfnamefont {R.~M.}\ \bibnamefont {Kerr}},\ and\
  \bibinfo {author} {\bibfnamefont {K.~R.}\ \bibnamefont {Sreenivasan}},\
  }\bibfield  {title} {\bibinfo {title} {Approach and separation of quantised
  vortices with balanced cores},\ }\href {https://doi.org/DOI:
  10.1017/jfm.2016.638} {\bibfield  {journal} {\bibinfo  {journal} {Journal of
  Fluid Mechanics}\ }\textbf {\bibinfo {volume} {808}},\ \bibinfo {pages} {641}
  (\bibinfo {year} {2016})}\BibitemShut {NoStop}%
\bibitem [{\citenamefont {Galantucci}\ \emph
  {et~al.}(2019{\natexlab{a}})\citenamefont {Galantucci}, \citenamefont
  {Baggaley}, \citenamefont {Parker},\ and\ \citenamefont
  {Barenghi}}]{Galantucci:2019aa}%
  \BibitemOpen
  \bibfield  {author} {\bibinfo {author} {\bibfnamefont {L.}~\bibnamefont
  {Galantucci}}, \bibinfo {author} {\bibfnamefont {A.~W.}\ \bibnamefont
  {Baggaley}}, \bibinfo {author} {\bibfnamefont {N.~G.}\ \bibnamefont
  {Parker}},\ and\ \bibinfo {author} {\bibfnamefont {C.~F.}\ \bibnamefont
  {Barenghi}},\ }\bibfield  {title} {\bibinfo {title} {Crossover from
  interaction to driven regimes in quantum vortex reconnections},\ }\href
  {https://doi.org/10.1073/pnas.1818668116} {\bibfield  {journal} {\bibinfo
  {journal} {Proceedings of the National Academy of Sciences}\ }\textbf
  {\bibinfo {volume} {116}},\ \bibinfo {pages} {12204} (\bibinfo {year}
  {2019}{\natexlab{a}})}\BibitemShut {NoStop}%
\bibitem [{\citenamefont {Rica}(2019)}]{Rica:2019aa}%
  \BibitemOpen
  \bibfield  {author} {\bibinfo {author} {\bibfnamefont {S.}~\bibnamefont
  {Rica}},\ }\bibfield  {title} {\bibinfo {title} {Self-similar vortex
  reconnection},\ }\bibfield  {booktitle} {\emph {\bibinfo {booktitle}
  {Patterns and dynamics: homage to Pierre Coullet / Formes et dynamique:
  hommage {\`a} Pierre Coullet}},\ }\href
  {https://doi.org/https://doi.org/10.1016/j.crme.2019.03.011} {\bibfield
  {journal} {\bibinfo  {journal} {Comptes Rendus M{\'e}canique}\ }\textbf
  {\bibinfo {volume} {347}},\ \bibinfo {pages} {365} (\bibinfo {year}
  {2019})}\BibitemShut {NoStop}%
\bibitem [{\citenamefont {Leadbeater}\ \emph {et~al.}(2001)\citenamefont
  {Leadbeater}, \citenamefont {Winiecki}, \citenamefont {Samuels},
  \citenamefont {Barenghi},\ and\ \citenamefont {Adams}}]{PhysRevLett.86.1410}%
  \BibitemOpen
  \bibfield  {author} {\bibinfo {author} {\bibfnamefont {M.}~\bibnamefont
  {Leadbeater}}, \bibinfo {author} {\bibfnamefont {T.}~\bibnamefont
  {Winiecki}}, \bibinfo {author} {\bibfnamefont {D.~C.}\ \bibnamefont
  {Samuels}}, \bibinfo {author} {\bibfnamefont {C.~F.}\ \bibnamefont
  {Barenghi}},\ and\ \bibinfo {author} {\bibfnamefont {C.~S.}\ \bibnamefont
  {Adams}},\ }\bibfield  {title} {\bibinfo {title} {Sound emission due to
  superfluid vortex reconnections},\ }\href
  {https://doi.org/10.1103/PhysRevLett.86.1410} {\bibfield  {journal} {\bibinfo
   {journal} {Phys. Rev. Lett.}\ }\textbf {\bibinfo {volume} {86}},\ \bibinfo
  {pages} {1410} (\bibinfo {year} {2001})}\BibitemShut {NoStop}%
\bibitem [{\citenamefont {Scheeler}\ \emph {et~al.}(2014)\citenamefont
  {Scheeler}, \citenamefont {Kleckner}, \citenamefont {Proment}, \citenamefont
  {Kindlmann},\ and\ \citenamefont {Irvine}}]{Scheeler28102014PNASDavide}%
  \BibitemOpen
  \bibfield  {author} {\bibinfo {author} {\bibfnamefont {M.~W.}\ \bibnamefont
  {Scheeler}}, \bibinfo {author} {\bibfnamefont {D.}~\bibnamefont {Kleckner}},
  \bibinfo {author} {\bibfnamefont {D.}~\bibnamefont {Proment}}, \bibinfo
  {author} {\bibfnamefont {G.~L.}\ \bibnamefont {Kindlmann}},\ and\ \bibinfo
  {author} {\bibfnamefont {W.~T.~M.}\ \bibnamefont {Irvine}},\ }\bibfield
  {title} {\bibinfo {title} {Helicity conservation by flow across scales in
  reconnecting vortex links and knots},\ }\href
  {https://doi.org/10.1073/pnas.1407232111} {\bibfield  {journal} {\bibinfo
  {journal} {Proceedings of the National Academy of Sciences}\ }\textbf
  {\bibinfo {volume} {111}},\ \bibinfo {pages} {15350} (\bibinfo {year}
  {2014})},\ \Eprint
  {https://arxiv.org/abs/http://www.pnas.org/content/111/43/15350.full.pdf}
  {http://www.pnas.org/content/111/43/15350.full.pdf} \BibitemShut {NoStop}%
\bibitem [{\citenamefont {Laing}\ \emph {et~al.}(2015)\citenamefont {Laing},
  \citenamefont {Ricca},\ and\ \citenamefont
  {De~Witt}}]{laing2015conservation}%
  \BibitemOpen
  \bibfield  {author} {\bibinfo {author} {\bibfnamefont {C.~E.}\ \bibnamefont
  {Laing}}, \bibinfo {author} {\bibfnamefont {R.~L.}\ \bibnamefont {Ricca}},\
  and\ \bibinfo {author} {\bibfnamefont {L.~S.}\ \bibnamefont {De~Witt}},\
  }\bibfield  {title} {\bibinfo {title} {Conservation of writhe helicity under
  anti-parallel reconnection},\ }\href@noop {} {\bibfield  {journal} {\bibinfo
  {journal} {Scientific reports}\ }\textbf {\bibinfo {volume} {5}} (\bibinfo
  {year} {2015})}\BibitemShut {NoStop}%
\bibitem [{\citenamefont {Clark~di Leoni}\ \emph {et~al.}(2016)\citenamefont
  {Clark~di Leoni}, \citenamefont {Mininni},\ and\ \citenamefont
  {Brachet}}]{diLeoniHelicity}%
  \BibitemOpen
  \bibfield  {author} {\bibinfo {author} {\bibfnamefont {P.}~\bibnamefont
  {Clark~di Leoni}}, \bibinfo {author} {\bibfnamefont {P.~D.}\ \bibnamefont
  {Mininni}},\ and\ \bibinfo {author} {\bibfnamefont {M.~E.}\ \bibnamefont
  {Brachet}},\ }\bibfield  {title} {\bibinfo {title} {Helicity, topology, and
  kelvin waves in reconnecting quantum knots},\ }\href
  {https://doi.org/10.1103/PhysRevA.94.043605} {\bibfield  {journal} {\bibinfo
  {journal} {Phys. Rev. A}\ }\textbf {\bibinfo {volume} {94}},\ \bibinfo
  {pages} {043605} (\bibinfo {year} {2016})}\BibitemShut {NoStop}%
\bibitem [{\citenamefont {Salman}(2017)}]{Salman:2017aa}%
  \BibitemOpen
  \bibfield  {author} {\bibinfo {author} {\bibfnamefont {H.}~\bibnamefont
  {Salman}},\ }\bibfield  {title} {\bibinfo {title} {Helicity conservation and
  twisted seifert surfaces for superfluid vortices},\ }\bibfield  {booktitle}
  {\emph {\bibinfo {booktitle} {Proceedings of the Royal Society A:
  Mathematical, Physical and Engineering Sciences}},\ }\href
  {https://doi.org/10.1098/rspa.2016.0853} {\bibfield  {journal} {\bibinfo
  {journal} {Proceedings of the Royal Society A: Mathematical, Physical and
  Engineering Sciences}\ }\textbf {\bibinfo {volume} {473}},\ \bibinfo {pages}
  {20160853} (\bibinfo {year} {2017})}\BibitemShut {NoStop}%
\bibitem [{\citenamefont {Zuccher}\ and\ \citenamefont
  {Ricca}(2015)}]{Zuccher&RiccaPRE2015}%
  \BibitemOpen
  \bibfield  {author} {\bibinfo {author} {\bibfnamefont {S.}~\bibnamefont
  {Zuccher}}\ and\ \bibinfo {author} {\bibfnamefont {R.~L.}\ \bibnamefont
  {Ricca}},\ }\bibfield  {title} {\bibinfo {title} {Helicity conservation under
  quantum reconnection of vortex rings},\ }\href
  {https://doi.org/10.1103/PhysRevE.92.061001} {\bibfield  {journal} {\bibinfo
  {journal} {Phys. Rev. E}\ }\textbf {\bibinfo {volume} {92}},\ \bibinfo
  {pages} {061001} (\bibinfo {year} {2015})}\BibitemShut {NoStop}%
\bibitem [{\citenamefont {Galantucci}\ \emph
  {et~al.}(2019{\natexlab{b}})\citenamefont {Galantucci}, \citenamefont
  {Baggaley}, \citenamefont {Parker},\ and\ \citenamefont
  {Barenghi}}]{galantucci2019crossover}%
  \BibitemOpen
  \bibfield  {author} {\bibinfo {author} {\bibfnamefont {L.}~\bibnamefont
  {Galantucci}}, \bibinfo {author} {\bibfnamefont {A.~W.}\ \bibnamefont
  {Baggaley}}, \bibinfo {author} {\bibfnamefont {N.~G.}\ \bibnamefont
  {Parker}},\ and\ \bibinfo {author} {\bibfnamefont {C.~F.}\ \bibnamefont
  {Barenghi}},\ }\bibfield  {title} {\bibinfo {title} {Crossover from
  interaction to driven regimes in quantum vortex reconnections},\ }\href@noop
  {} {\bibfield  {journal} {\bibinfo  {journal} {Proceedings of the National
  Academy of Sciences}\ }\textbf {\bibinfo {volume} {116}},\ \bibinfo {pages}
  {12204} (\bibinfo {year} {2019}{\natexlab{b}})}\BibitemShut {NoStop}%
\bibitem [{\citenamefont {Proment}\ and\ \citenamefont
  {Krstulovic}(2020)}]{Proment2020MatchingSupplemental}%
  \BibitemOpen
  \bibfield  {author} {\bibinfo {author} {\bibfnamefont {D.}~\bibnamefont
  {Proment}}\ and\ \bibinfo {author} {\bibfnamefont {G.}~\bibnamefont
  {Krstulovic}},\ }\bibfield  {title} {\bibinfo {title} {Supplemental material
  for: A matching theory to characterize sound emission during vortex
  reconnection in quantum fluids},\ }\href@noop {} {\bibfield  {journal}
  {\bibinfo  {journal} {Mathematica Notebook available as Supplemental
  Material}\ } (\bibinfo {year} {2020})}\BibitemShut {NoStop}%
\bibitem [{\citenamefont {Kerr}(2011)}]{KerrPRL2011}%
  \BibitemOpen
  \bibfield  {author} {\bibinfo {author} {\bibfnamefont {R.~M.}\ \bibnamefont
  {Kerr}},\ }\bibfield  {title} {\bibinfo {title} {Vortex stretching as a
  mechanism for quantum kinetic energy decay},\ }\href
  {https://doi.org/10.1103/PhysRevLett.106.224501} {\bibfield  {journal}
  {\bibinfo  {journal} {Phys. Rev. Lett.}\ }\textbf {\bibinfo {volume} {106}},\
  \bibinfo {pages} {224501} (\bibinfo {year} {2011})}\BibitemShut {NoStop}%
\bibitem [{\citenamefont {Pismen}\ and\ \citenamefont
  {Pismen}(1999)}]{Pismen:1999aa}%
  \BibitemOpen
  \bibfield  {author} {\bibinfo {author} {\bibfnamefont {L.~M.}\ \bibnamefont
  {Pismen}}\ and\ \bibinfo {author} {\bibfnamefont {L.~M.}\ \bibnamefont
  {Pismen}},\ }\href@noop {} {\emph {\bibinfo {title} {Vortices in nonlinear
  fields: from liquid crystals to superfluids, from non-equilibrium patterns to
  cosmic strings}}},\ Vol.\ \bibinfo {volume} {100}\ (\bibinfo  {publisher}
  {Oxford University Press},\ \bibinfo {year} {1999})\BibitemShut {NoStop}%
\bibitem [{\citenamefont
  {Sonin}(1987)}]{soninVortexOscillationsHydrodynamics1987}%
  \BibitemOpen
  \bibfield  {author} {\bibinfo {author} {\bibfnamefont {E.~B.}\ \bibnamefont
  {Sonin}},\ }\bibfield  {title} {\bibinfo {title} {Vortex oscillations and
  hydrodynamics of rotating superfluids},\ }\href
  {https://doi.org/10.1103/RevModPhys.59.87} {\bibfield  {journal} {\bibinfo
  {journal} {Reviews of Modern Physics}\ }\textbf {\bibinfo {volume} {59}},\
  \bibinfo {pages} {87} (\bibinfo {year} {1987})}\BibitemShut {NoStop}%
\bibitem [{\citenamefont {Jones}\ and\ \citenamefont
  {Roberts}(1982)}]{Jones:1982aa}%
  \BibitemOpen
  \bibfield  {author} {\bibinfo {author} {\bibfnamefont {C.~A.}\ \bibnamefont
  {Jones}}\ and\ \bibinfo {author} {\bibfnamefont {P.~H.}\ \bibnamefont
  {Roberts}},\ }\bibfield  {title} {\bibinfo {title} {Motions in a bose
  condensate. iv. axisymmetric solitary waves},\ }\href
  {https://doi.org/10.1088/0305-4470/15/8/036} {\bibfield  {journal} {\bibinfo
  {journal} {Journal of Physics A: Mathematical and General}\ }\textbf
  {\bibinfo {volume} {15}},\ \bibinfo {pages} {2599} (\bibinfo {year}
  {1982})}\BibitemShut {NoStop}%
\bibitem [{\citenamefont {Svancara}\ and\ \citenamefont
  {La~Mantia}(2019)}]{svancaralamantia2019}%
  \BibitemOpen
  \bibfield  {author} {\bibinfo {author} {\bibfnamefont {P.}~\bibnamefont
  {Svancara}}\ and\ \bibinfo {author} {\bibfnamefont {M.}~\bibnamefont
  {La~Mantia}},\ }\bibfield  {title} {\bibinfo {title} {Flight-crash events in
  superfluid turbulence},\ }\href {https://doi.org/10.1017/jfm.2019.586}
  {\bibfield  {journal} {\bibinfo  {journal} {Journal of Fluid Mechanics}\
  }\textbf {\bibinfo {volume} {876}},\ \bibinfo {pages} {R2} (\bibinfo {year}
  {2019})}\BibitemShut {NoStop}%
\bibitem [{\citenamefont {Giuriato}\ and\ \citenamefont
  {Krstulovic}(2020)}]{GiuriatoReconnections}%
  \BibitemOpen
  \bibfield  {author} {\bibinfo {author} {\bibfnamefont {U.}~\bibnamefont
  {Giuriato}}\ and\ \bibinfo {author} {\bibfnamefont {G.}~\bibnamefont
  {Krstulovic}},\ }\bibfield  {title} {\bibinfo {title} {Quantum vortex
  reconnections mediated by trapped particles},\ }\href
  {https://doi.org/10.1103/PhysRevB.102.094508} {\bibfield  {journal} {\bibinfo
   {journal} {Phys. Rev. B}\ }\textbf {\bibinfo {volume} {102}},\ \bibinfo
  {pages} {094508} (\bibinfo {year} {2020})}\BibitemShut {NoStop}%
\bibitem [{\citenamefont {Frisch}(1995)}]{Frisch1995}%
  \BibitemOpen
  \bibfield  {author} {\bibinfo {author} {\bibfnamefont {U.}~\bibnamefont
  {Frisch}},\ }\href@noop {} {\emph {\bibinfo {title} {{Turbulence: The Legacy
  of A. N. Kolmogorov}}}}\ (\bibinfo  {publisher} {Cambridge University
  Press},\ \bibinfo {year} {1995})\BibitemShut {NoStop}%
\bibitem [{\citenamefont {Serafini}\ \emph {et~al.}(2015)\citenamefont
  {Serafini}, \citenamefont {Barbiero}, \citenamefont {Debortoli},
  \citenamefont {Donadello}, \citenamefont {Larcher}, \citenamefont {Dalfovo},
  \citenamefont {Lamporesi},\ and\ \citenamefont
  {Ferrari}}]{PhysRevLett.115.170402}%
  \BibitemOpen
  \bibfield  {author} {\bibinfo {author} {\bibfnamefont {S.}~\bibnamefont
  {Serafini}}, \bibinfo {author} {\bibfnamefont {M.}~\bibnamefont {Barbiero}},
  \bibinfo {author} {\bibfnamefont {M.}~\bibnamefont {Debortoli}}, \bibinfo
  {author} {\bibfnamefont {S.}~\bibnamefont {Donadello}}, \bibinfo {author}
  {\bibfnamefont {F.}~\bibnamefont {Larcher}}, \bibinfo {author} {\bibfnamefont
  {F.}~\bibnamefont {Dalfovo}}, \bibinfo {author} {\bibfnamefont
  {G.}~\bibnamefont {Lamporesi}},\ and\ \bibinfo {author} {\bibfnamefont
  {G.}~\bibnamefont {Ferrari}},\ }\bibfield  {title} {\bibinfo {title}
  {Dynamics and interaction of vortex lines in an elongated bose-einstein
  condensate},\ }\href {https://doi.org/10.1103/PhysRevLett.115.170402}
  {\bibfield  {journal} {\bibinfo  {journal} {Phys. Rev. Lett.}\ }\textbf
  {\bibinfo {volume} {115}},\ \bibinfo {pages} {170402} (\bibinfo {year}
  {2015})}\BibitemShut {NoStop}%
\bibitem [{\citenamefont {Xhani}\ \emph {et~al.}(2020)\citenamefont {Xhani},
  \citenamefont {Neri}, \citenamefont {Galantucci}, \citenamefont {Scazza},
  \citenamefont {Burchianti}, \citenamefont {Lee}, \citenamefont {Barenghi},
  \citenamefont {Trombettoni}, \citenamefont {Inguscio}, \citenamefont
  {Zaccanti}, \citenamefont {Roati},\ and\ \citenamefont
  {Proukakis}}]{Xhani:2020aa}%
  \BibitemOpen
  \bibfield  {author} {\bibinfo {author} {\bibfnamefont {K.}~\bibnamefont
  {Xhani}}, \bibinfo {author} {\bibfnamefont {E.}~\bibnamefont {Neri}},
  \bibinfo {author} {\bibfnamefont {L.}~\bibnamefont {Galantucci}}, \bibinfo
  {author} {\bibfnamefont {F.}~\bibnamefont {Scazza}}, \bibinfo {author}
  {\bibfnamefont {A.}~\bibnamefont {Burchianti}}, \bibinfo {author}
  {\bibfnamefont {K.~L.}\ \bibnamefont {Lee}}, \bibinfo {author} {\bibfnamefont
  {C.~F.}\ \bibnamefont {Barenghi}}, \bibinfo {author} {\bibfnamefont
  {A.}~\bibnamefont {Trombettoni}}, \bibinfo {author} {\bibfnamefont
  {M.}~\bibnamefont {Inguscio}}, \bibinfo {author} {\bibfnamefont
  {M.}~\bibnamefont {Zaccanti}}, \bibinfo {author} {\bibfnamefont
  {G.}~\bibnamefont {Roati}},\ and\ \bibinfo {author} {\bibfnamefont {N.~P.}\
  \bibnamefont {Proukakis}},\ }\bibfield  {title} {\bibinfo {title} {Critical
  transport and vortex dynamics in a thin atomic josephson junction},\ }\href
  {https://doi.org/10.1103/PhysRevLett.124.045301} {\bibfield  {journal}
  {\bibinfo  {journal} {Physical Review Letters}\ }\textbf {\bibinfo {volume}
  {124}},\ \bibinfo {pages} {045301} (\bibinfo {year} {2020})}\BibitemShut
  {NoStop}%
\bibitem [{\citenamefont {Amelio}\ \emph {et~al.}(2018)\citenamefont {Amelio},
  \citenamefont {Galli},\ and\ \citenamefont
  {Reatto}}]{PhysRevLett.121.015302}%
  \BibitemOpen
  \bibfield  {author} {\bibinfo {author} {\bibfnamefont {I.}~\bibnamefont
  {Amelio}}, \bibinfo {author} {\bibfnamefont {D.~E.}\ \bibnamefont {Galli}},\
  and\ \bibinfo {author} {\bibfnamefont {L.}~\bibnamefont {Reatto}},\
  }\bibfield  {title} {\bibinfo {title} {Probing quantum turbulence in
  $^{4}\mathrm{He}$ by quantum evaporation measurements},\ }\href
  {https://doi.org/10.1103/PhysRevLett.121.015302} {\bibfield  {journal}
  {\bibinfo  {journal} {Phys. Rev. Lett.}\ }\textbf {\bibinfo {volume} {121}},\
  \bibinfo {pages} {015302} (\bibinfo {year} {2018})}\BibitemShut {NoStop}%
\bibitem [{\citenamefont {Proment}\ \emph {et~al.}(2012)\citenamefont
  {Proment}, \citenamefont {Onorato},\ and\ \citenamefont
  {Barenghi}}]{PhysRevE.85.036306}%
  \BibitemOpen
  \bibfield  {author} {\bibinfo {author} {\bibfnamefont {D.}~\bibnamefont
  {Proment}}, \bibinfo {author} {\bibfnamefont {M.}~\bibnamefont {Onorato}},\
  and\ \bibinfo {author} {\bibfnamefont {C.~F.}\ \bibnamefont {Barenghi}},\
  }\bibfield  {title} {\bibinfo {title} {Vortex knots in a bose-einstein
  condensate},\ }\href {https://doi.org/10.1103/PhysRevE.85.036306} {\bibfield
  {journal} {\bibinfo  {journal} {Phys. Rev. E}\ }\textbf {\bibinfo {volume}
  {85}},\ \bibinfo {pages} {036306} (\bibinfo {year} {2012})}\BibitemShut
  {NoStop}%
\bibitem [{\citenamefont {Kleckner}\ \emph {et~al.}(2016)\citenamefont
  {Kleckner}, \citenamefont {Kauffman},\ and\ \citenamefont
  {Irvine}}]{Kleckner:2016aa}%
  \BibitemOpen
  \bibfield  {author} {\bibinfo {author} {\bibfnamefont {D.}~\bibnamefont
  {Kleckner}}, \bibinfo {author} {\bibfnamefont {L.~H.}\ \bibnamefont
  {Kauffman}},\ and\ \bibinfo {author} {\bibfnamefont {W.~T.~M.}\ \bibnamefont
  {Irvine}},\ }\bibfield  {title} {\bibinfo {title} {How superfluid vortex
  knots untie},\ }\href {https://doi.org/10.1038/nphys3679} {\bibfield
  {journal} {\bibinfo  {journal} {Nature Physics}\ }\textbf {\bibinfo {volume}
  {12}},\ \bibinfo {pages} {650 EP } (\bibinfo {year} {2016})}\BibitemShut
  {NoStop}%
\bibitem [{\citenamefont {Alekseenko}\ \emph {et~al.}(2016)\citenamefont
  {Alekseenko}, \citenamefont {Kuibin}, \citenamefont {Shtork}, \citenamefont
  {Skripkin},\ and\ \citenamefont {Tsoy}}]{Alekseenko:2016aa}%
  \BibitemOpen
  \bibfield  {author} {\bibinfo {author} {\bibfnamefont {S.~V.}\ \bibnamefont
  {Alekseenko}}, \bibinfo {author} {\bibfnamefont {P.~A.}\ \bibnamefont
  {Kuibin}}, \bibinfo {author} {\bibfnamefont {S.~I.}\ \bibnamefont {Shtork}},
  \bibinfo {author} {\bibfnamefont {S.~G.}\ \bibnamefont {Skripkin}},\ and\
  \bibinfo {author} {\bibfnamefont {M.~A.}\ \bibnamefont {Tsoy}},\ }\bibfield
  {title} {\bibinfo {title} {Vortex reconnection in a swirling flow},\
  }\href@noop {} {\bibfield  {journal} {\bibinfo  {journal} {JETP Lett.}\
  }\textbf {\bibinfo {volume} {103}},\ \bibinfo {pages} {455} (\bibinfo {year}
  {2016})}\BibitemShut {NoStop}%
\end{thebibliography}

%

\end{document}